\newcount\draft\draft=1 
\newcount\cameraready\cameraready=0

\documentclass[letterpaper]{sig-alternate} 

\usepackage{mathptmx} 

\newcommand{\ignore}[1]{}
\usepackage{fancyhdr}
\usepackage[hyphens]{url}
\usepackage{hyperref}
\usepackage{graphicx}
\usepackage{xspace}
\usepackage{subfig}
\usepackage{makecell}
\DeclareGraphicsExtensions{.pdf,.jpg,.png}
\graphicspath{{./Figures} {./figs/} {./plots/}}

\newcommand{\sys}{CCache\xspace}
\newcommand{\cdata}{CData\xspace}
\newcommand{\cop}{COp\xspace}
\newcommand{\cops}{COps\xspace}
\newcommand{\cgl}{CGL\xspace}
\newcommand{\fgl}{FGL\xspace}

\newcommand{\dup}{DUP\xspace}
\newcommand{\cread}{CRead\xspace}
\newcommand{\cwrite}{CWrite\xspace}
\newcommand{\gups}{Key-value Store\xspace}
\newcommand{\CREAD}{\texttt{c\_read}\xspace}
\newcommand{\CWRITE}{\texttt{c\_write}\xspace}
\newcommand{\RDMR}{\texttt{rd\_mreg}\xspace}
\newcommand{\WRMR}{\texttt{wr\_mreg}\xspace}
\newcommand{\smrg}{\texttt{soft\_merge}\xspace}
\newcommand{\hmrg}{\texttt{merge}\xspace}
\begin{document}
\title{Flexible Support for Fast Parallel Commutative Updates}
\author{
  Vignesh Balaji \qquad Dhruva Tirumala \qquad Brandon Lucia \\
  \\
  Carnegie Mellon University \\
  \\
  \{vigneshb, dtirumal, blucia\}@andrew.cmu.edu \\
  }


\maketitle

\begin{abstract}
Privatizing data is a useful strategy for increasing parallelism in a shared
memory multithreaded program.  Independent cores can compute independently on
duplicates of shared data, combining their results at the end of their
computations. Conventional approaches to privatization, however, rely on explicit
static or dynamic memory allocation for duplicated state, increasing memory
footprint and contention for cache resources, especially in shared caches.  In
this work, we describe CCache, a system for on-demand privatization of data
manipulated by commutative operations. CCache garners the benefits of
privatization, without the increase in memory footprint or cache
occupancy.  Each core in CCache dynamically privatizes commutatively
manipulated data, operating on a copy.  Periodically or
at the end of its computation, the core merges its value with the value
resident in memory, and when all cores have merged, the in-memory copy contains
the up-to-date value.  We describe a low-complexity architectural
implementation of CCache that extends a conventional multicore to support
on-demand privatization without using additional memory for private copies. We
evaluate CCache on several high-value applications, including random access
key-value store, clustering, breadth first search and graph ranking,
showing speedups upto 3.2X.
\end{abstract}

\section{Introduction}
As parallel computers and programming languages have proliferated, programmers
are increasingly faced with the task of improving software performance through
parallelism.  Shared-memory multithreading is an especially common programming
and execution model that is at the heart of a wide variety of server and user
applications.  The value of the shared memory model is its simple programming
interface.  However, shared memory also requires programmers to overcome
several barriers to realize an application's parallel performance potential.

{\em Synchronization} and {\em data movement} are the key impediments to an
application's efficient, parallel execution.  To ensure that data shared by
multiple threads remain consistent, the programmer must use synchronization
(e.g., mutex locks~\cite{pthreads}) to serialize the threads' accesses to the
data.  Synchronization limits parallelism because it forces threads to
sequentially access shared resources, often requiring threads to stop executing
and wait for one another.  Processors' data caches are essential to high
performance, and the need to manipulate shared data in cache requires the
system to move the data between different processors' caches during an
execution.  The latency of data movement impedes performance.  Moreover,
systems must use cache coherence~\cite{coherenceprimer} to ensure that
processors always operate on the most up-to-date version of a value.  Coherence
protocol implementations cause processors to serialize their accesses to shared
data, further limiting parallelism and performance.

Our work is motivated by an observation about synchronization and
data movement: while accesses to shared data by different threads {\em must be
serialized}, the order in which those accesses are serialized is often
inconsequential.  Indeed, in a multithreaded execution, the execution order of
such accesses may vary non-deterministically, potentially leading to different
-- yet correct -- outcomes.  We refer to operations with this permissible form
of order non-determinism as ``commutative operations'' (or COps) and the data
that they access as ``commutatively accessible data'' (or CData).  

Recent work described COUP~\cite{coup}, which modified the coherence protocol
to exploit the commutativity of common operations (e.g., addition, logical OR).
While COUP is effective at improving parallelism for programs that use these
commutative operations, COUP has several important limitations.  COUP is
limited to a small, fixed set of operations that are built into the hardware.
If software uses even a slightly different commutative updates (e.g.,
saturating addition, complex arithmetic) COUP is inapplicable and its
performance benefits are lost.  Additionally, COUP tightly couples commutative
updates to the coherence protocol, adding a new coherence state, along with its
attendant complexity and need for re-verification.

This work describes a hybrid software/hardware approach to exploiting the
commutativity of COps on CData. We describe \sys, which uses simple hardware
support that does not modify the cache coherence protocol to improve the
parallel performance of threads executing flexible, software-defined
commutative operations.  Cores in \sys perform COps to replicated, {\em
privatized copies} of the same CData without the need for synchronization,
coherence, or data movement. Threads that perform parallel COps on replicated
CData must eventually {\em merge} the result of their COps using an
application-specific merge function that the programmer writes, to combine
independently manipulated copies of CData.  Merging combines the CData results
of different threads' COps, effectively serializing the execution of the
parallel COps.  \sys improves parallel performance through {\em on-demand
privatization}, creating a copy of CData on which each thread may perform COps
independently.

We describe extensions to a commodity multicore architecture that support \sys.
\sys's microarchitectural additions have low complexity and do not interfere
with critical path operations. \sys uses a simple set of ISA extensions to
support a programming interface that allows programmers to express \cops and
define, register, and execute a merge function.  \sys requires commutatively
manipulated data and coherently manipulated data to be disjoint, enabling
efficient commutative updates with coherence protocol modifications. Coherent
cache lines are handled by the existing coherence protocol.  Commutatively
manipulated lines never generate coherence actions and never match the tag of
an incoming coherence message.  \sys thus avoids the cost and complexity 
of a protocol change.

Section~\ref{sec:eval} evaluates \sys on a collection of benchmarks including a
key-value store, K-means clustering, Breadth-first Search~\cite{gap} and
PageRank~\cite{pagerank}. To illustrate the flexibility of \sys, we implement
variants of each benchmark that use different, application-specific merge
operations.  Using direct comparisons to static data duplication and
fine-grained locking, our evaluation shows \sys improves the single-machine,
in-memory performance of these applications by upto 3.2x over an already
optimized, parallel baseline.  Moreover, with {\em half the
L3 cache capacity}, \sys has a 1.07-1.9x performance improvement over static
duplication.

To summarize, our main contributions are:
\begin{itemize}
\item The \sys execution model, which uses on-demand privatization 
to improve parallel performance of commutative shared data accesses.
\item We present a collection of architecture extensions that implement
on-demand privatization in \sys without affecting the coherence protocol.
\item We port several important applications to use \sys's ISA extensions,
including a key-value store, PageRank, BFS, and K-means clustering.
\item We implement static duplication and fine-grained locking implementations of
our workloads, and we show by direct comparison that \sys improves performance
upto 3.2X across applications.
\end{itemize}

\section{Background and Motivation}
\label{sec:background} 
This section motivates the \sys approach to on-demand privatization in
hardware.  We frame \sys with a discussion of fine-grained locking and
static data duplication, done manually~\cite{revisionsOOPSLA10,revisionsOOPSLA11} or
with compiler support~\cite{openmpreduce}.

\subsection{Locking and Data Duplication}

\begin{figure}[h]
\includegraphics[width=0.9\columnwidth]{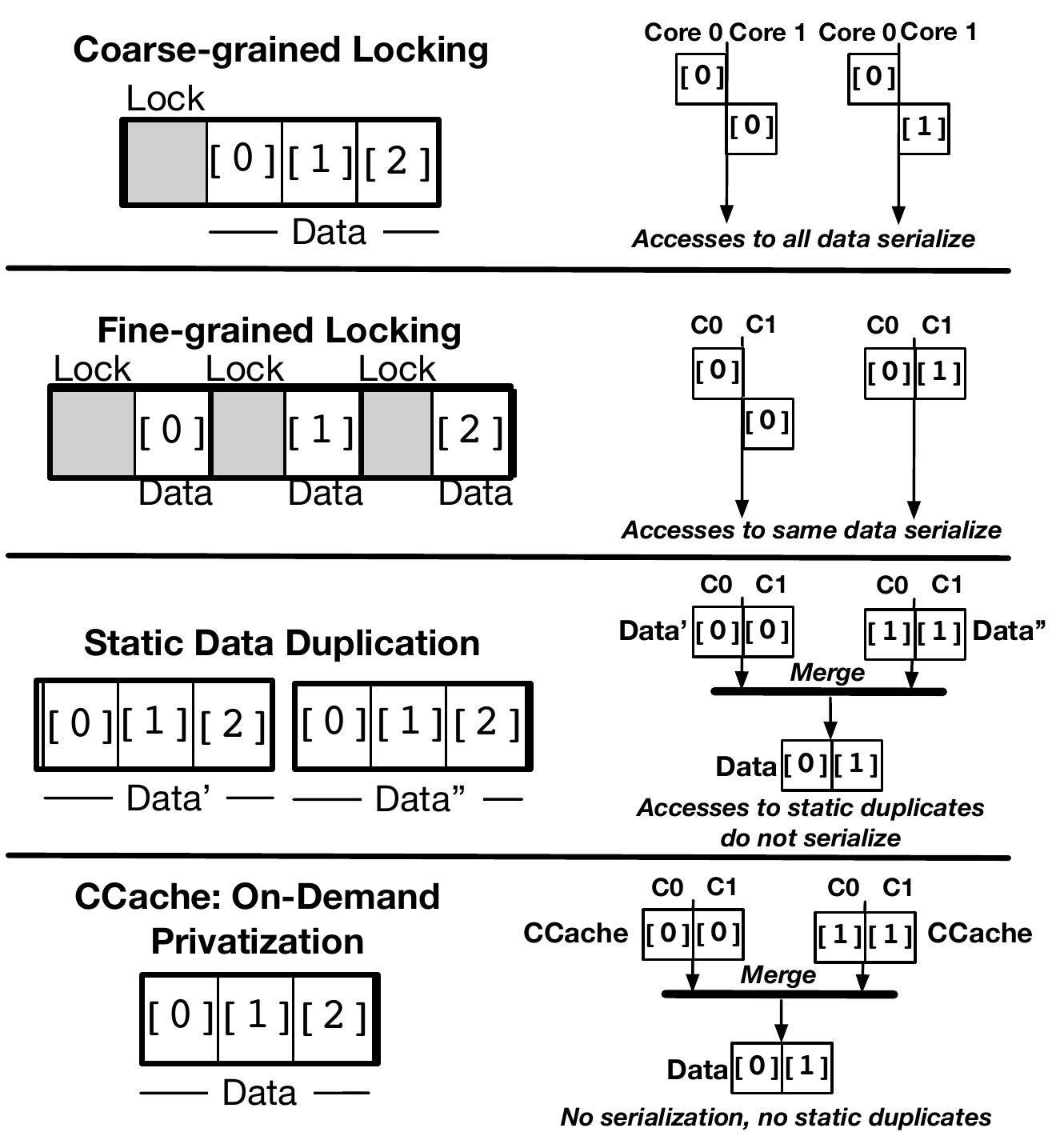}
\caption{\label{fig:cglfgldup}{\bf Locking, Data Duplication, and \sys.} \cgl
permits little parallelism.   \fgl accesses are parallel for different
locations.  Accesses to duplicates (DUP) of a single location are parallel.
\fgl and \dup incur space overhead for locks/copies.  Parallel updates to
duplicates must be merged (not shown). \sys allows parallel access to all
locations without space overhead.}
\end{figure}

Parallel code requires threads to synchronize their accesses to shared data to
keep the data consistent.  {\em Lock-based synchronization} requires threads to
use {\em locks} associated with shared data to serialize the threads'
accesses to the shared data.  The simplest way to implement locking in a
parallel program is to use {\em coarse-grained locking} (\cgl).  \cgl
associates one lock (or a small number of locks) with a large, shared data
structures.  \cgl makes programming simple because the programmer is not
required to reason about the details of associating locks with each variable or
memory location.  However, \cgl can impede performance by serializing accesses
to unrelated memory locations that are protected by the same lock.  {\em
Fine-grained locking} (\fgl) is one response to the performance impediment of
\cgl.  \fgl associates a lock with each (or few) variables, eliminating
unnecessary serialization of accesses to unrelated data.  The key problem with
\fgl is the need for a programmer to express the mapping of locks to data,
which is more complex for \fgl than for \cgl, and is a source of errors.
Figure~\ref{fig:cglfgldup} illustrates the difference between \fgl and \cgl.

\subsubsection{Data Duplication}
{\em Data duplication} (\dup) is a strategy for increasing parallelism by creating
copies of a memory location that different threads can manipulate independently
of one another.   To ensure the correctness of an execution that both reads
{\em and} writes duplicated data, the program must, at some point, combine the
results computed by the different threads.  Functional
reductions~\cite{openmpreduce,mapreduceCACM08,mapreduceOSDI04} and symbolic
parallel updates~\cite{commutativityanalysis,sympleSOSP15,retconISCA10} combine
the result of each thread's computation on its duplicated copy of the data.
Reduction applies a (usually) side-effect-free operation to all of the copies
producing a single, coherent output value.  Some prior work has statically
replicated data using compiler and runtime support, parceling copies out to
threads and merging them with a
reduction~\cite{openmpreduce,mapreduceCACM08,mapreduceOSDI04,sympleSOSP15}.
Other prior work has used hardware support for speculation to
effectively duplicate data~\cite{stmTOCS07,retconISCA10}.

\dup is highly effective, especially when threads can independently
perform many operations on their copy of the data.  Duplication improves
parallel performance by eliminating serialization due to synchronization (i.e.,
locking) and cache coherence, both of which hinder \fgl and \cgl. Instead, the
reduction computes a result that is equivalent to some serialization of the
independent computations.

Despite its benefits, \dup has several drawbacks.  Duplicating data
increases the application's memory footprint.  The increase in footprint leads
to higher LLC occupancy and miss rate.  Efficiently laying out and distributing
duplicated data is difficult; we discuss this programming difficulty of data
duplication in the context of our benchmarks in Section~\ref{sec:setup}.
Anecdotally, we found static data duplication more difficult to get right than \fgl,
forcing us to simultaneously reason about consistency, locality, and false
sharing.

Our main insight is that data duplication and locking both have merits and
drawbacks: data duplication increases parallelism at a cost in memory and LLC
misses; locking decreases parallelism and makes programming complex, but does
not degrade LLC performance or increase the memory footprint.  Our work aims to
capture the ``best of all worlds'': the complexity and occupancy in the LLC and
memory of \cgl (or \fgl), and the increased parallelism of \dup.  As
Figure~\ref{fig:cglfgldup} shows, \sys, our novel programming model and
architecture, dynamically privatizes and later merges data without the need for
the programmer to tediously lay out and manage in-memory copies.
Additionally, our mechanism can use the \emph{same} merge function as defined
for \dup. Hence, \sys applies generally, to \emph{all} cases where static \dup
is possible.  Section~\ref{sec:idea} describes the new \sys approach at a high
level and Section~\ref{sec:arch} describes our programming model and
architecture.

\section{\sys: On-Demand Privatization}
\label{sec:idea}
\sys is a new programming and execution model for a parallel program that uses {\em on-demand}
data duplication to increase parallel access to data (\cdata) that are
manipulated by commutative operations (\cops).   When several cores access the
same \cdata memory locations using \cops, the data are {\em privatized}, providing a
copy to each core.  After privatization, the cores can manipulate the
\cdata in their caches in parallel without coherence or synchronization.  When
a core finishes operating on \cdata location, it uses a programmer-defined {\em
merge function} to merge its updated value back into the multiprocessor's
closest level of shared storage (i.e., the LLC or memory).  When all cores have
merged their privatized copies, the result is the equivalent to some
serialization of the cores' parallel manipulations of the \cdata.  

We describe the operation of \sys, assuming a baseline multicore
architecture with an arbitrary cache hierarchy.  In Section~\ref{sec:arch} we
describe a concrete architectural incarnation of \sys.  There are two parts to
\sys.   We first define \cops and \cdata and show how
on-demand data duplication increases parallelism.  Then we describe what \sys
does when a core executes a \cop to a \cdata memory location.  Last, we
describe what a \sys core does when its commutative \cdata computation
completes.

\subsection{Executing Commutative Operations in Parallel}

\begin{figure}[tbp] \centering
\includegraphics[width=\columnwidth]{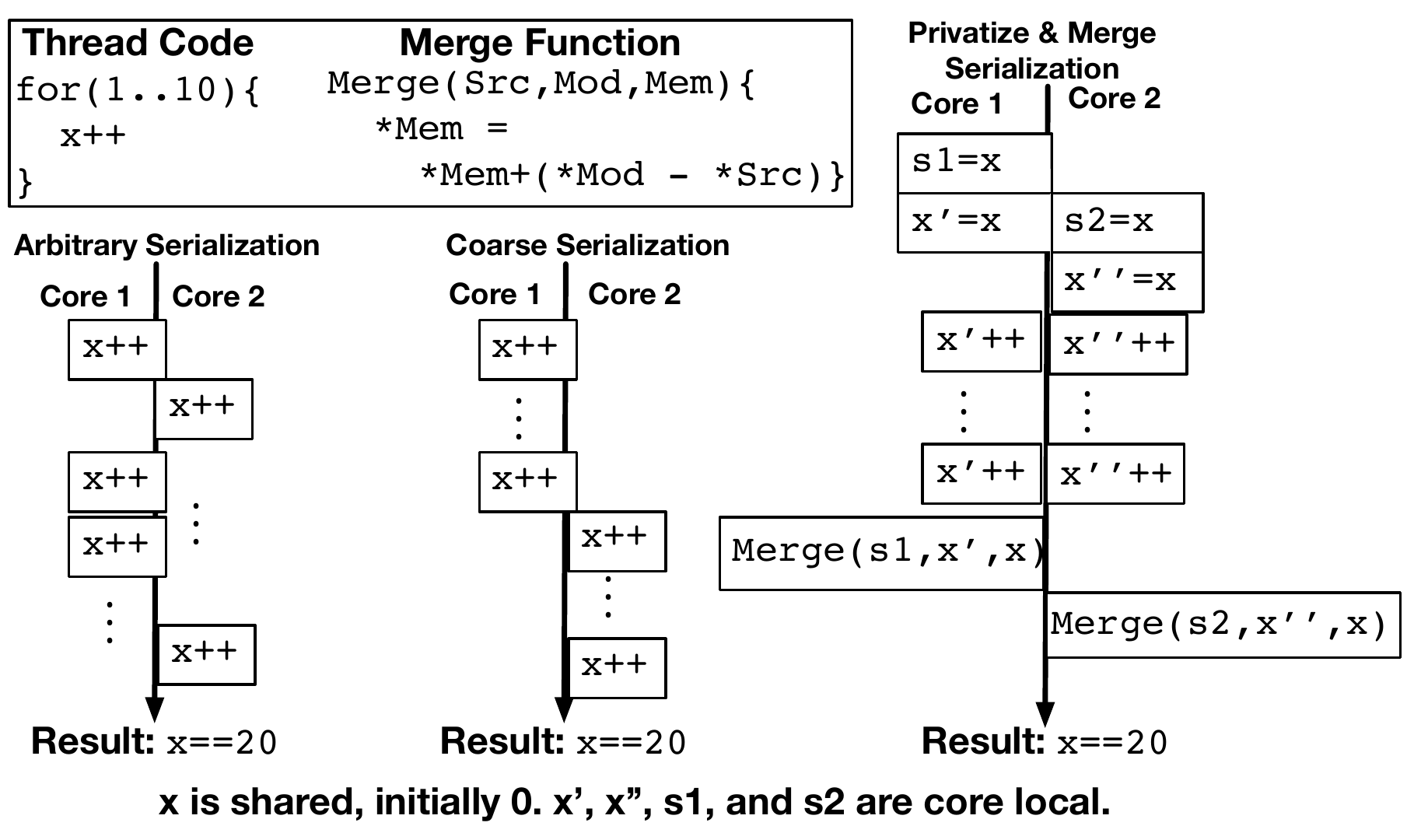}
\caption{\label{fig:mergingoverview} {\bf Three Ways to Serialize Commutative Updates.} }
\end{figure}

\sys increases the parallel performance of a program's {\em commutative}
manipulations of shared data.  
Operations to a shared memory location by different cores are commutative with
one another if their execution can be {\em serialized} in either possible
execution order and produce a correct result.  Figure~\ref{fig:mergingoverview} shows a
simple program in which two cores increment a shared counter variable {\tt x}.
The figure illustrates that these operations are commutative -- the arbitrary
serialization of the loop's iterations and the coarse serialization produce the
same result at the end of the execution. Any other serialization of the updates
to {\tt x} yield the same result.

\sys increases a core's parallel performance running operations that access
memory commutatively (\cops) by automatically, dynamically duplicating the
memory being accessed.  \sys requires the programmer to explicitly identify
which memory operations in the program access data commutatively (\cops).  The
\cops defined in \sys are the \cread and the \cwrite operations.    A \cread or
a \cwrite operation creates two copies of memory location it is accessing, the
first is the core-local {\em source copy}, which the core preserves.  The
second copy is the core-local {\em update copy}, which the core uses to perform
its computation, instead of referring directly to the location in {\em memory}.
Each core executes its \cops independently on a privatized copy
of the shared \cdata, and then {\em merges} the resulting privatized copies,
producing a result that is equivalent to some serialization.  We discuss
merging in Section~\ref{subsec:merge_hl}.

Figure~\ref{fig:mergingoverview} shows how duplication improves parallelism in
the \sys-like ``privatize \& merge'' serialization depicted.  The two cores
privatize the value of {\tt x} by preserving its source value into the
abstract storage locations {\tt s1} and {\tt s2} and copying 
{\tt x} into core-local, abstract storage locations {\tt x$'$} and {\tt x$''$}.
The cores then independently execute their loops, updating their private
copies.  Note that this ``privatize \& merge'' execution model for manipulating
commutative data does not specify {\em where} to put the abstract storage for
the copies.  To simplify our exposition, we show data copies in named
variables in the figure, but \sys {\em  does not use explicit, named
copies}.  As Section~\ref{sec:arch:rw} describes, \sys stores the updated
copy in the core's private cache, and its source copy in a new hardware
structure called the {\em source buffer}.  Architecture support for 
privatizing data is crucial, avoiding the memory overhead of statically
allocated copies and the time overhead of dynamically allocating copies in
software.

\subsection{Merging Updates to Privatized Data}\label{subsec:merge_hl}

A {\em merge function} in \sys is a programmer-provided, application-specific
function that uses a core's updated value, saved source value, and the
in-memory value to update the in-memory copy.  Merging is a partial
reduction~\cite{openmpreduce,mapreduceOSDI04,mapreduceCACM08} of a core's
value and the in-memory copy.


A typical merge function examines the {\em difference} between the source copy
and the update copy to compute the {\em update} to apply to the memory copy.
The merge function then applies the update to the memory copy, reflecting the
execution of the core's \cops.  When a set of cores that are commutatively
manipulating data have all applied their merge functions, the data are
consistent and represent a serialization of the cores' updates.   

The flexibility of a software-defined merge function is one of the most
important, novel aspects of \sys, allowing its applicability to a broad variety
of applications, and allowing applications (and their merge functions) to
evolve with time.  

Possible software merge functions in \sys include, {\em but are not limited to}
complex multiplication, saturating or thresholding addition, arbitrary bitwise
logic, and variants using floating- and fixed-point operations. \sys also
supports approximate merge techniques. An example of an approximate merge is to
dynamically, selectively drop updates according to a programmer-provided
binomial distribution, similar to loop perforation~\cite{loopperforation}.
Each of these merge functions is represented in a benchmark that we use to
evaluate \sys in Section~\ref{sec:eval}.  We emphasize that \sys is a stark
contrast to a system with fixed, hardware merge operations~\cite{coup}, which
are less broadly applicable and unable to evolve to changing application needs.  

Figure~\ref{fig:mergingoverview} shows how merging produces a correct
serialized result.  After a core complete its update loop, it executes the
programmer-provided merge function shown.  The programmer writes the merge
function with the knowledge of the updates applied by the threads in their
loops -- the execution applies a sequence of increments to {\tt x}.    The
merge function computes the update to apply to memory. In this example, the
update is to add to the memory value the difference between the core's modified
value and the preserved source value.  To apply the update, the merge function
{\em adds} the computed difference to the value in memory.  After both cores
execute their merge function, {\tt x} is in a consistent final state,
equivalent to both the arbitrary and the coarse-grained serialization.

\subsubsection{Synchronization and Merging}

Parallel merges to the same location must be serialized for correctness and the
execution of each merge to a location must be atomic with respect to that
location. Such per-merge synchronization ensures that each subsequent merge
sees the updated memory result of previous merges ensuring the result is a
serialization of atomically applied updates.  Section~\ref{sec:arch} describes
how our \sys architecture serializes merges.

In addition to serialized, atomic merges, the programmer may sometimes need a
barrier-like {\em merge boundary} that pauses every thread manipulating \cdata
until all threads have merged all \cdata.  A program needs a merge boundary
when it is transitioning between phases and the next phase needs results from
the previous phase.  A programmer can implement a merge boundary by executing a
standard barrier, preceded by an explicit \sys merge operation in each core
that merges all duplicated values.  If each core executes the \sys merge
operation and then waits at the barrier, the merge boundary leaves all data
consistent.  Note that when such a barrier is needed in \sys, it would also be
needed in a conventional program and \sys imposes no additional need for
barrier synchronization.

\subsection{Example: A \sys Key-value Store}
\begin{figure}[tbp]  \centering
\includegraphics[width=\columnwidth]{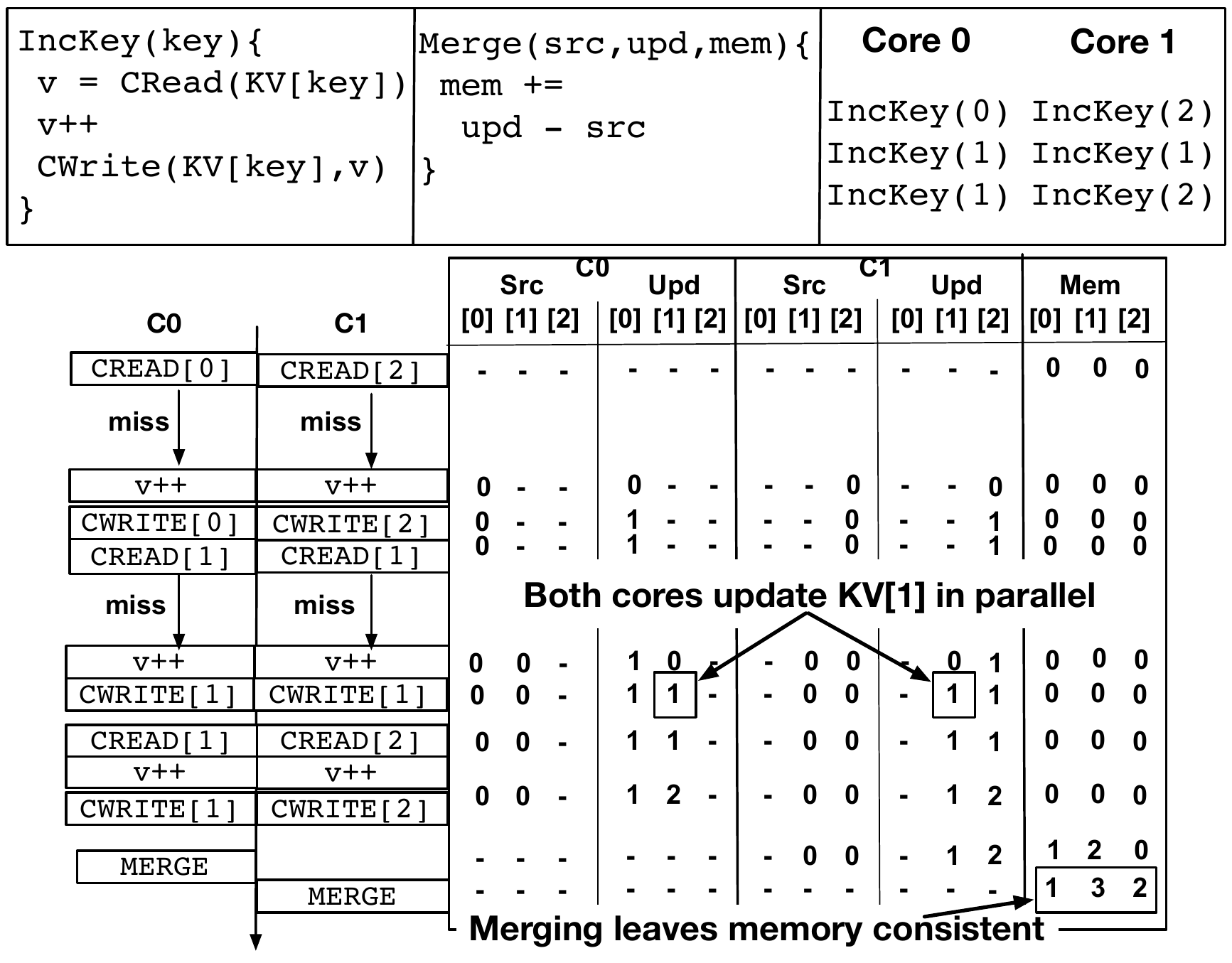}
\caption{\label{fig:kv_store} {\bf A Key-value Store Executing with \sys.} 
Cores execute the key updates shown.  Time goes top to bottom for the execution
at left and state updates at right.  The state at right shows the cores' source
copies, {\em Src}, the cores' updated copies, {\em Upd}, and the in-memory
copy, {\em Mem}.  Data are {\em not} statically duplicated; cores 
privatize them on-demand. The merge step applies
the merge function to all entries in {\tt KV} (not shown).}
\end{figure}

We illustrate how \sys's on-demand data duplication idea increases parallelism
with an example.  Figure~\ref{fig:kv_store} shows a key-value store manipulated
by two cores.  The program keeps a lookup table {\tt KV} indexed by key,
containing integer values that the cores increment.  We use \cread and \cwrite
operations that perform on-demand data duplication.  The merge function takes
the source value at the time it was first read and privatized by the \cread,
the updated value, and the shared memory value.  The merge function adds the
difference between the updated value and the source value to the memory value.

The figure reveals several important characteristics of \sys.  First, the
figure shows how \sys obviates duplicating {\tt KV}.
Instead, cores copy individual entries of {\tt KV} {\em on-demand} into
the Src and Upd copies.   Second, the example shows that by privatizing {\tt
KV[1]}, the cores can independently read {\em and} write its value in
parallel.  Third, the figure shows that core 1 has {\em locality} in its
independent accesses to its privatized copy of {\tt KV[1]}.  Fourth, the
figure shows that the serialized execution of the merge functions by each core
installs correct, consistent final values into shared memory.

\section{\sys Architecture}\label{sec:arch}
We co-designed \sys as a collection of programming primitives implemented in a
commodity multicore architecture.  We assume a base multicore architecture with
core-private L1 and L2 caches and a shared last-level cache (LLC) with a
directory-based, MESI cache coherence protocol. 

\begin{figure}[h]
\centering
  \includegraphics[keepaspectratio,width=.9\columnwidth]{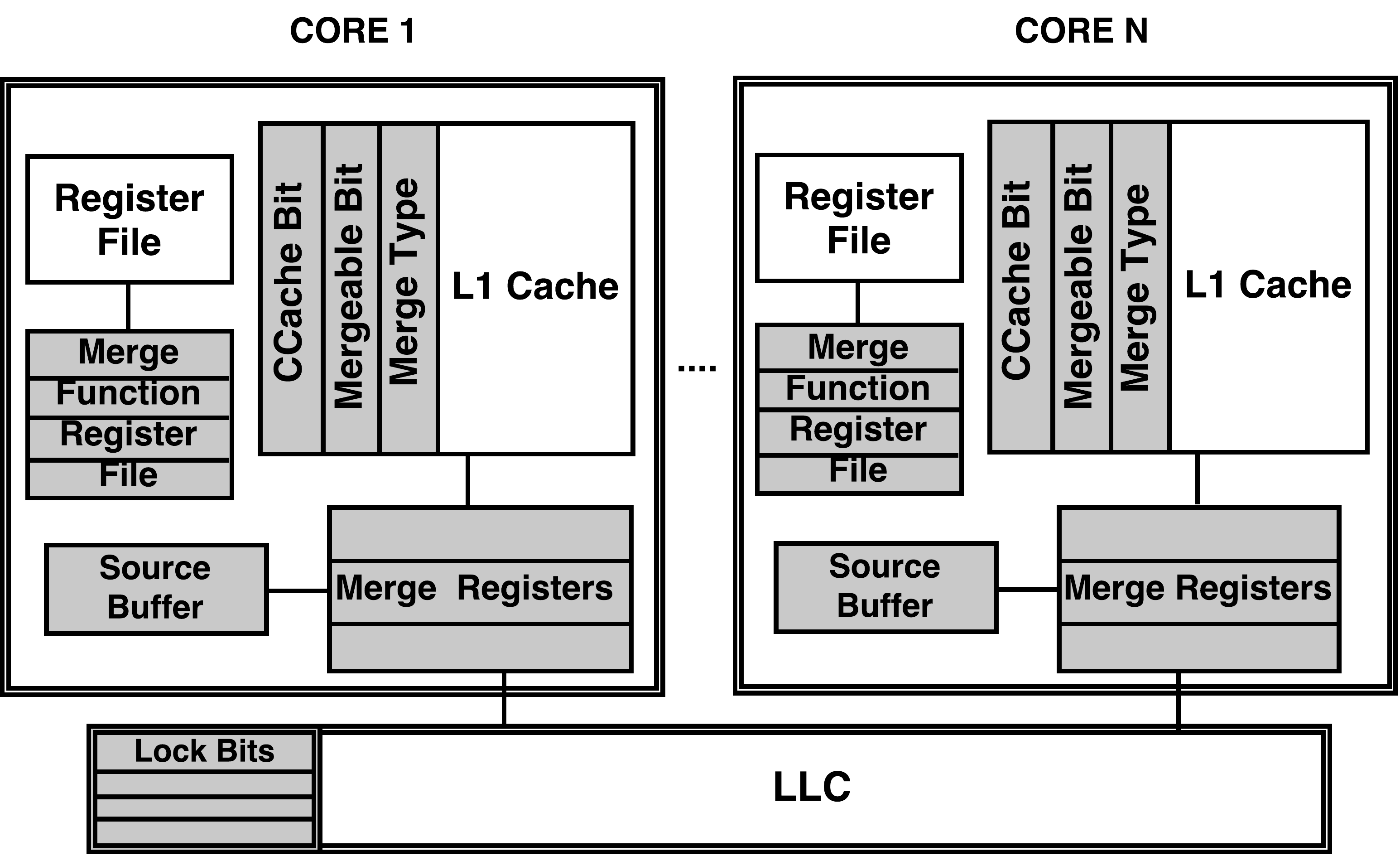}
  \caption{\textbf{The architectural modifications for \sys}}
  \label{fig:arch}
\end{figure}

The \sys programming interface includes operations for manipulating and merging
\cdata, which are summarized in Table~\ref{table:ISA}. 
We implement the \sys programming primitives directly as ISA extensions using
modifications to the L1 cache and a dedicated {\em source buffer} that manages
source copies of \cdata.
We add support for \sys maintain a collection of merge functions and associate
each \cdata line with its merge function. 
Figure~\ref{fig:arch} shows the structures that we add to our base architecture
design.   
Beyond the basic \sys design, we improve the performance of merging with an
optimization that improves locality and eliminates unnecessary evictions.

\begin{table}[t]
\scriptsize
\renewcommand{\tabcolsep}{2pt}
\begin{tabular}{l|l}
\Xhline{2pt}
\textbf{\sys Primitive} & \textbf{Operation} \\
\Xhline{1pt}
\texttt{merge\_init(\&fn,i)} & Stores pointer to merge function {\tt fn} into MFR {\tt i} \\
\hline
\texttt{c\_read(\cdata,i)} & Read \cdata into src. buff. \& L1, set \sys bit,\\
                           & set merge type to {\tt i} \\
\hline
\texttt{c\_write(\cdata,v,i)} & Read \cdata into src. buff. on miss, write v in L1,\\
                              & set \sys bit if unset, set merge type to {\tt i}\\
\hline
\texttt{rd\_mreg(reg,i)} & Return word {\tt i} of merge register {\tt reg}  \\
\hline
\texttt{wr\_mreg(reg,v,i)} & Write {\tt v} into word {\tt i} of merge register {\tt reg} \\
\hline
\texttt{soft\_merge(core)} & Set mergeable bit in L1 for each valid source buffer entry \\
\hline
\texttt{merge(core\_id)} & For each valid source buffer entry, populate merge registers, \\
                         & lock LLC line, call line's merge function from MFRF, copy \\
                         & memory merge register to LLC, flash clear source buffer, \\
			 & unset \sys bit, unlock LLC line. \\
\end{tabular}
\caption{\label{table:ISA}{\bf \sys programming primitives.}}

\end{table}

\subsection{\cread and \cwrite}
\label{sec:arch:rw}
We introduce \CREAD and \CWRITE operations that access \cdata in a similar way
to typical load and store instructions.  When a core executes a \CREAD or
\CWRITE operation, it loads the data into its L1 cache, as usual, but does not
perform any coherence actions for the involved line.  The core also need not
acquire any lock before accessing \cdata with a \CREAD or \CWRITE.

To track which lines hold \cdata, we add a single {\em \sys bit} to each line
in the L1 cache.  When a \CREAD or a \CWRITE accesses a line for the first time
(i.e., on an L1 miss), the core sets the line's \sys bit.  We also add a field
to each cache line that describes the {\em merge type} of the data in the line.
The merge type field determines which merge function to call when merging a
line's \cdata (Section~\ref{subsec:merge}).  The size of the merge type field
is the logarithm of the number of different merge functions in the system.  An
implementation using two bits (i.e., four merge functions) is reasonably
flexible and makes the merge type field's hardware overhead very low.

To allow for update-based merging, \sys must maintain the {\em source copy},
{\em updated copy}, and {\em memory copy}, as described in
Section~\ref{subsec:merge_hl}.  \sys uses the L1 cache itself to maintain the updated
copy and keeps the memory copy in shared memory as usual. 

To maintain the {\em source copy} of a line accessed by a \CREAD or \CWRITE we
add a dedicated hardware structure to each core called the {\em source buffer}.
The source buffer is a small, fully associative, cache-like memory that stores
data at the granularity of cache lines.  Figure~\ref{fig:arch} illustrates
the source buffer in the context of the entire core.  When a \CREAD or \CWRITE
experiences an L1 miss, \sys copies the value into an entry in the source
buffer in parallel with loading the line into the L1.

\subsection{Merging}\label{subsec:merge}

The programmer {\em registers} a programmer-defined merge function for a region
of \cdata using \sys's \texttt{merge\_init} operation.  At a merge point, the
system executes the merge function, passing as arguments the memory, source,
and updated value of the \cdata location to be merged. The result of a merge
function is that the memory copy of the data reflects the updates executed by
the merging core before the merge.  The signature of a merge function is fixed.
A merge function takes pointers to three 64-byte values: the source and updated
values are read-only inputs and the memory value acts as both an input and an
output.  The merge function must read and write these values using the \RDMR
and \WRMR \sys operations depicted in Table~\ref{table:ISA}.

{\noindent \bf Registering Merge Functions}
To implement merging, we add a {\em merge function register file} (MFRF) to the
architecture to hold the addresses of merge function.   We add a
\texttt{merge\_init} ISA instruction that installs a merge function pointer
into a specified MFRF entry.  The size of the MFRF is dictated by number of
{\em simultaneous} \cdata types in the system.  A four entry MFRF allows four
different merge types and requires only two merge type bits per cache line to
identify a line's merge function.

{\noindent \bf Executing a Merge Function}
\sys runs a merge function at a {\em merge operation}.  Table~\ref{table:ISA}
shows \sys's two varieties of merge - \smrg and \hmrg.  We discuss the basic
\hmrg instruction here and defer discussion of the optimized \smrg to
Section~\ref{subsec:optimizations}.  

A \hmrg merges all of a core's \cdata: the executing core walks the source
buffer array and executes a merge for each valid entry.  To execute a merge for
a line, the core first locks the corresponding line in the LLC, preventing any
other core from accessing the line until the merge completes.  To prevent
deadlock, a merge function can access only its source, updated, and memory
values.  Allowing arbitrary access to LLC data could cause two threads in merge
functions to wait on one anothers' locked LLC lines.

\begin{figure}[h] \centering
\includegraphics[width=\columnwidth]{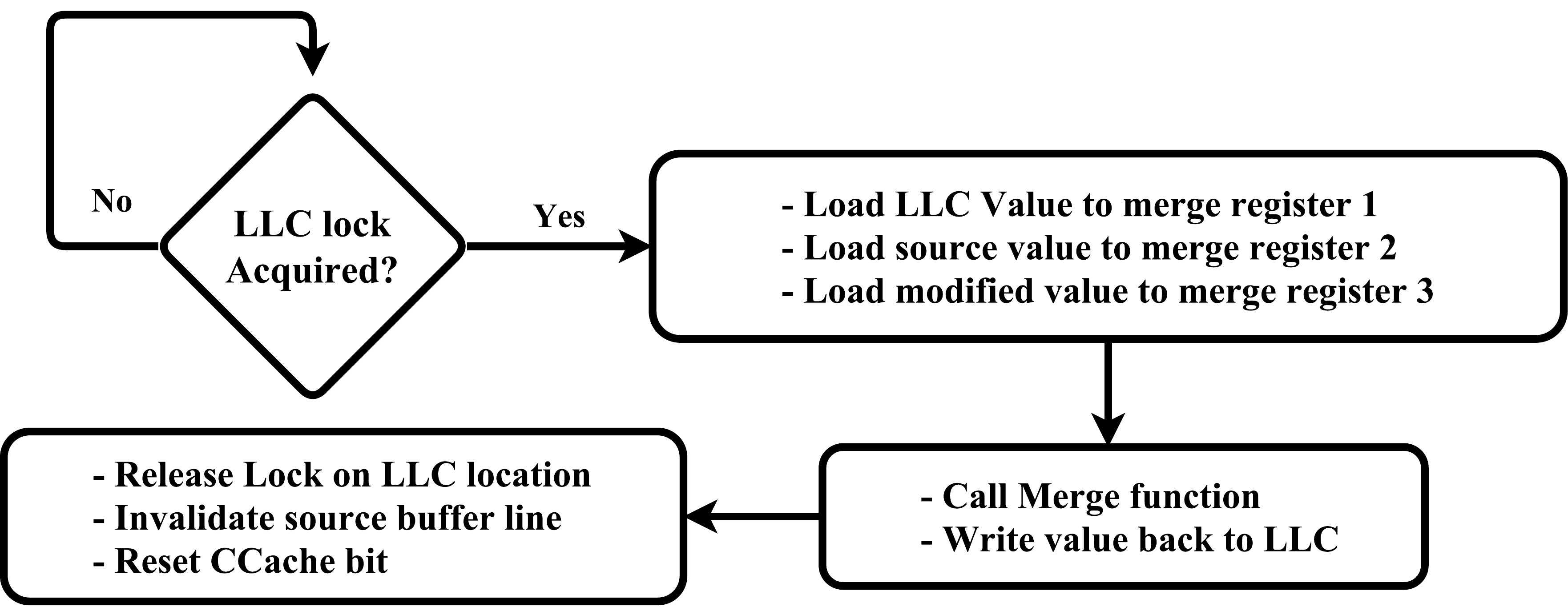}
\caption{\label{fig:flow}{\bf Merging a Source Buffer Entry.}}
\end{figure}

After locking the LLC line, the core next prepares the source, updated, and
memory values for the merge.   To prepare them, the core copies the value of
each into its own, dedicated, cache-line-sized {\em merge registers}, that we
add to the core (see Figure~\ref{fig:arch}).    After preparing the merge
registers, the core calls the merge function and as it executes, \RDMR and
\WRMR \sys operations that refer to the memory, updated, and source copies of
\cdata access the copies in the merge registers.  After the merge function
completes, the core moves the contents of the merge register that corresponds
to the memory copy of the merged line and into the L1 and triggers a write back
to the LLC. Additionally, the source buffer line is invalidated and the \sys
bit is reset to zero. The core then unlocks the LLC line, completing the merge.
The entire sequence of steps during merging is shown in the flowchart in
Figure~\ref{fig:flow}.

{\noindent \bf Serialization and Merge Functions.} A \hmrg instruction
serializes accesses to each merged LLC line by individually locking and
unlocking them.  The \hmrg does {\em not} enforce the atomicity of a group of
merge operations to different lines in the source buffer.  Only individual
lines' merges are atomic.  For coarser atomicity, the programmer should use
locks and barriers as usual.  Note that any situation in \sys that calls for a
lock or barrier would require at least the same synchronization (or more) in a
non-\sys program because such a point requires results to be serialized,
regardless of the programming model.

\subsection{Merge Optimizations}\label{subsec:optimizations}

\hmrg operations incur the cost of merging all of a core's \cdata, including
merge function execution and write back to the LLC.
We applied two optimizations to \sys to reduce the cost of \hmrg operations. 
First, we introduce a new instruction, \smrg, that delays the merge and write
back of \cdata lines for as long as possible. Second, we perform the \hmrg 
operation described above only when a core updates a \cdata line.

\smrg works by setting a \cdata line into a new {\em mergeable} state
and delaying the merge of the line's contents with the in-memory copy until 
the line's eviction from the L1 cache and source buffer. To track lines in 
the mergeable state, we add a new {\em mergeable} bit per cache line in the 
L1, which is depicted in Figure~\ref{fig:arch}. A \smrg operation sets a 
line's mergeable bit, indicating that it is ready to merge. When a core must evict from L1, lines with their
\sys and mergeable bits are candidates for eviction. (Recall that lines
with only their \sys bit set cannot be evicted.) When a mergeable line is 
selected for eviction, the core first executes the merge procedure (from
Section~\ref{subsec:merge}) for the line and then evicts it. A \CREAD or 
\CWRITE to a mergeable line resets the line's mergeable bit to
prevent the line from being evicted during subsequent commutative updates.  These
\CREAD or \CWRITE operations enjoy additional locality in the source
buffer and the L1 cache, compared to our unoptimized implementation.

The second optimization is to not perform \hmrg operations
for clean \cdata lines because merging an unmodified copy would not affect the in-memory result. \sys checks a line's L1 dirty bit 
to decide whether a \hmrg operation is required. \cdata lines which are candidates
for eviction (ie. mergeable bit set) and do not have their dirty bit set can be 
silently evicted from the L1 cache and removed from the source buffer. 

%
%

\subsection{Correctness}\label{subsec:correctness}

\sys does not require modifications to the cache coherence protocol, avoiding a
major source of complexity and verification cost.  The cache coherence protocol
operates as usual for non-commutatively, coherently manipulated lines.  \sys
does not require sending any new coherence messages because a core never
generates a coherence messages for a line with its \sys bit set.  \sys also
does not require a core to specially handle any incoming coherence message
because no incoming coherence message can ever refer to a line of \cdata. A
coherence messages cannot refer to a \cdata line because a \cdata line can only
ever be manipulated by a \cread or \cwrite operation with the line's \sys bit,
which never generates any coherence messages.

\sys affects the cache's replacement policy because \cdata are not
allowed to be evicted.  \sys must ensure that data accessed without coherence
by a \CREAD or \CWRITE are merged before being evicted.  However, \sys
cannot simply merge on an eviction because words from the line might be
modified in registers.  If such a line were evicted along with its source
buffer entry, then when the register value was written back using a \CWRITE
operation, the source buffer entry would no longer be available.  Furthermore,
the in-memory value may have changed (due to writes by other cores) by the time
of the \CWRITE.  The absence of a source value and potential for changes to the
memory copy in this situation precludes the eviction, and \sys conservatively
prohibits all evictions of data with their \sys bit set.  

A program cannot access more cache lines using \cops than there are ways in the
cache without an intervening merge.  If there are $w$ ways in the cache, \sys
will deadlock after $w+1$ \cops if all accessed data map to the same cache set.
Consequently, the programmer needs to carefully ensure that their program
accesses at most $w-1$ distinct cache lines without an intervening merge. A
limit of $w-1$ guarantees that in the worst case, when all accesses map to the
same cache set, one way in the set is always available for coherent data,
access to which may be required to make progress.  In systems with SMT,
hardware threads evenly divide cache ways for \cdata.  While somewhat limiting,
this programming restriction is similar to recent, wide-spread hardware
transactional memory proposals~\cite{tsx,asf}.

We assume \cdata are cache line aligned and that lines containing \cdata are
only ever accessed using a \CREAD or \CWRITE instruction.  We require the
programmer or the compiler to add padding bytes to these aligned \cdata
variables to ensure that a cache line never contains both \cdata and normal
data bytes.  This restriction prevents operations other than \CREAD and \CWRITE
operations from accessing \cdata lines.  

\subsection{Commutativity of the Merge Function}
\label{sec:conditional}
In \sys, the programmer is solely responsible for the correctness of merge
functions.   The key to writing a merge function is to determine what update to
apply to the in-memory copy of the cache line, given the updated copy, and
source buffer copy.  Merge functions are often arithmetic and 
computing and applying the update is simple.  We have written many
such cases (e.g., addition, minimum) that can be used as a library with little
programmer effort.  

A modestly more complex case is a commutative update with a conditional that
depends on the values of the \cdata.  In this case, the programmer must ensure
that the merge function's conditional observes the in-memory copy of the value,
rather than the updated copy.  An example of such a program might randomly
access and increment an array of integers up to a maximum.  A simple merge for
integer addition adds the difference between the source value and the updated
value to the in-memory value.  To enforce a maximum, the merge function must
also assess whether applying the update would exceed the maximum and, if so,
update the in-memory copy to the maximum only.

\subsection{Handling Context Switches and Interrupts}
\sys cannot evict \cdata from the cache without merging.  However, at an
arbitrarily timed context switch or interrupt, a program may be using \cdata
(i.e., in registers), making a merge impossible.  There are two options for
handling these events.  The first option is to delay the context switch or
interrupt until after a \smrg executes for each \cdata line.  At such a point,
the architecture can safely execute the merge function for each \cdata line and
then execute the context switch as usual.  The main drawback to this approach
is the need to delay context switches and interrupts, which may not be possible
or desirable in some cases.  Additionally, the delay is unpredictable,
depending on the number of \cdata lines and the timing of merge operations.  

An alternative is to save cached \cdata and source buffer entries with the
process control block before the context switch.  When switching back, a
process must restore its \cdata and source buffer entries.  This approach
eliminates the delay of waiting to merge, but increases the size of process
state.  With an 8-way L1 and an 8-entry source buffer a process must preserve
at most 1KB of state --- a managable overhead for infrequent and already-costly
context switches.

\subsection{Area and Energy Overheads}
We used CACTI~\cite{cacti} to quantify the overhead in extending the 
microarchitecture for \sys. We observed that the energy and area overhead
of adding tracking bits to each cache line in the L1 cache and LLC would be
negligible. A 32 entry, fully associative source buffer would occupy ~0.1\%
the area of the Last Level Cache. The energy of reading and writing data 
from a source buffer of this size would be ~6.5\% of the energy of accessing
the LLC. We assumed a 32nm process for all the caches in the system.   

\section{EXPERIMENTAL SETUP}
\label{sec:setup}
In this section we describe the simulation setup we used to evaluate \sys.  We
built our simulation infrastructure using PIN~\cite{pin}.  To measure baseline
performance, we developed a simulator that modeled a 3-level cache hierarchy
implementing a directory-based MESI coherence protocol. To measure the
performance of \sys, we extended the baseline simulator code to model
\sys's architectural additions. Our
\sys simulator modeled incoherent accesses to \cdata, an 8-entry source
buffer and a modified cache replacement policy that excludes \cdata. We
also modeled the cost of executing merge functions in software, including the
cost of accessing the merge registers and the LLC.  Our model does not include
the latency incurred due to waiting on locked LLC lines, but concurrent merges of
the same line are rare and this simplification is unlikely to significantly
alter our results.  Table \eqref{table:cost} describes the architectural
parameters of our simulated architectures.   

\begin{table}[h] \centering
\scriptsize
\begin{tabular}{ l | l}
\Xhline{2pt}
\textbf{Processor} & 8-cores. Non-memory instructions: 1 cycle \\
\hline
\textbf{L1 cache} & 8-way, 32KB, 64B lines, 4 cyc./hit \\
\hline
\textbf{L2 cache} & 8-way, 512KB, 64B lines, 10 cyc./hit \\
\hline
\textbf{Shared LLC} & 16-way, 4MB, 64B lines, 70 cyc/hit \\
\hline
\textbf{Main memory} & 300 cyc./access \\
%
\Xhline{2pt}
\textbf{Source buffer} & fully assoc. 512B per-core, 64B lines, 3 cyc./hit \\
\hline
\textbf{Merge Latency} & 170 cycles incl. LLC round-trip \\
\end{tabular}
\caption{\label{table:cost}{\bf Simulated Architecture Parameters.}}
\end{table}


  
\subsection{Benchmarks}

To evaluate \sys, we manually ported four parallel applications that
commutatively manipulate shared data to use \sys:  a \gups, K-Means clustering
, Page Rank and BFS.  For each benchmark, we also implemented two variants: one that
uses fine-grained locks to protect data and the other statically duplicates
data. The following subsections provides a brief overview of
the operation of each application.

{\noindent \bf Key-Value Store}
A key-value store is a lookup table that associates a key with a value,
allowing a core to refer to and manipulate the value using the key.  In our
\gups benchmark, 8 cores increment the values associated with randomly chosen
keys.   We used \cops to implement the updates because increments commute. We
set the total number of accesses to random keys to 16 times the number of keys,
which we varied in our experiments from 250,000 to 4,000,000.  Our \dup 
scheme creates a per-thread copy of the value array.
The merge function computes the difference between the updated
copy and the source copy and adds the difference to the memory copy. We 
use the same merge function for both \sys and \dup. 

{\noindent \bf K-Means}
K-Means is a popular clustering algorithm.
The algorithm assigns a set of $n$ $m$-dimensional vectors into
$k$ clusters with the objective of minimizing the sum of distances between 
each point and its cluster's center. The algorithm iteratively assigns each 
point to the nearest cluster and then recomputes the cluster centers. To restrict
simulation times, we fix the number of iterations of the algorithm.

Our \dup implementation is based on Rodinia's~\cite{rodinia} static data 
duplication scheme, which creates a per-thread copy the cluster center data
structure. For the \sys implementation, we made the cluster centers \cdata 
and used \cops to manipulate them. The merge function for both \sys and \dup
does a component-wise addition of weights on point vectors in a cluster.

The results for K-Means also highlights the need for our soft-merge optimization
(described in section~\ref{subsec:optimizations}). The cluster centres stored in \sys 
experience high reuse over the course of the 
application. While a naive implementation of \sys would require the CData
to be merged after every iteration, the soft-merge optimization can
exploit the locality in CData by delaying the merge operation until \sys becomes full.
We discuss the benefit of this optimization in further detail in Section~\ref{subsec:soft-merge} 

{\noindent \bf Page Rank}
Page Rank~\cite{pagerank} is a relevance-based graph node ranking algorithm
that assigns a rank to a node based on its indegree and outdegree.  As the
algorithm computes the rank recursively, the data structure that contains each
node's rank is shared among all the threads.  A naive \dup implementation would
allocate each thread a private copy of the \emph{entire} node rank data
structure. Instead, we wrote an optimized data duplication implementation that
partitions nodes across threads and creates only one duplicate. One
copy of the structure holds the current iteration's updates while the other
copy is a read-only copy of the result of the previous iteration. These copies are 
then switched at the end of every iteration. For the \sys
version, we made the node rank data structure \cdata and manipulated it using
\cops.  To test Page Rank on varied inputs, we used three inputs generated by
the Graph500~\cite{graph500} benchmark input generator using the RMAT, SSCA,
and Random configurations.  The merge function adds an iteration's update
to the global rank.

{\noindent \bf Breath First Search (BFS)}
Breadth First Search is a graph traversal order that forms an important
kernel in many graph applications. For our evaluations, we
used the BFS kernel of the Betweeness Centrality application from the 
GAP benchmark suite~\cite{gap}. The implementation uses a bitmap
to efficiently represent the edges linking successors from a 
source vertex. The original implementation uses a Compare-and-Swap
to atomically set an entry in the bitmap. We replaced the atomic
operations with fine grained locks that matched the update granularity
of the set operation in our \fgl version. For the \dup version, we
used an optimization that avoids privatization of the entire bitmap.
Instead, we store all the updates from a thread in a thread-local
dynamically sized container and apply these updates atomically 
during a merge operation at the end. For the \sys version, we 
simply marked the bitmap as \cdata and used \cops to set locations
of the bitmap. The \sys merge function performs a logical OR of 
all the privatized copies. We evaluated the four versions on 
kronecker and uniform random graphs provided in the GAP benchmark.



{\noindent \bf Duplication Strategies}
Porting code from fine-grained locking to a memory-efficient  \dup version is a 
non-trivial task. We used an optimized \dup strategy for Page Rank because it
was inefficient to replicate the entire rank array across all threads. Similarly,
we avoid naive replication in BFS because the size of the bitmap makes creating
thread-local copies infeasible. By contrast, in K-Means, we observed that replicating the 
data structure storing cluster centers did not drastically increase memory 
footprint. As a testament to the complexity of efficient duplication, we found
that Rodinia's K-Means implementation suffered from high false sharing. 
The \gups imposes an application-level constraint
on duplication: partitioning is not a good match because any core
may access any key.  In our experiments, it was reasonable to duplicate the
table across all cores.  In general, making decisions about how to duplicate
data requires difficult, application-dependent reasoning.  \sys eliminates the
need for such subtle reasoning about data duplication, instead, just requiring
the programmer to use \cops. 


\section{Evaluation}\label{sec:eval}
\begin{figure*}[ht]
\begin{minipage}{\textwidth}
\begin{tabular}{cccc}
\subfloat[K-Means (FP)]{\label{kmeans-fp}\includegraphics[keepaspectratio,width=0.24\textwidth]{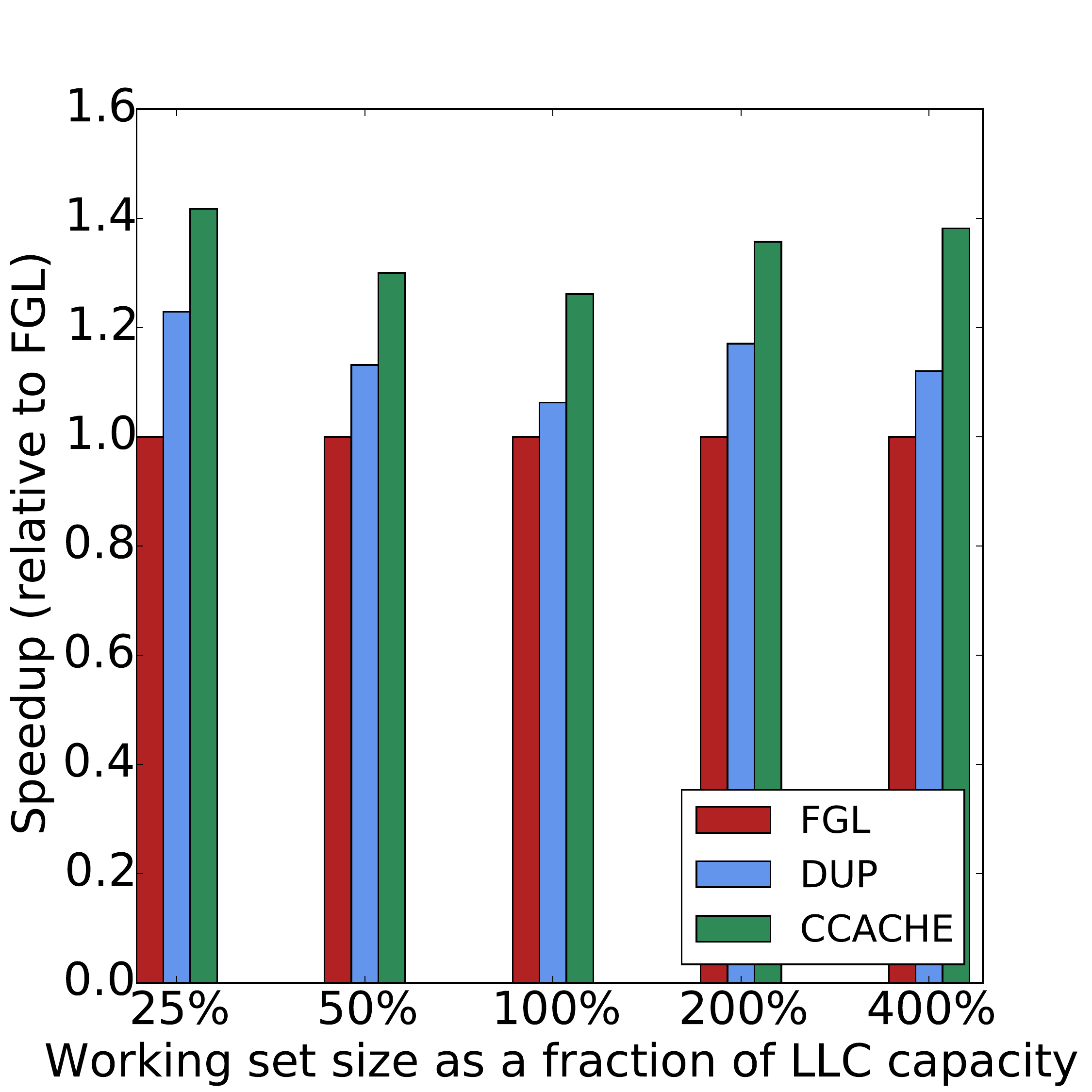}}
   & \subfloat[K-Means (int)]{\label{kmeans-int}\includegraphics[keepaspectratio,width=0.24\textwidth]{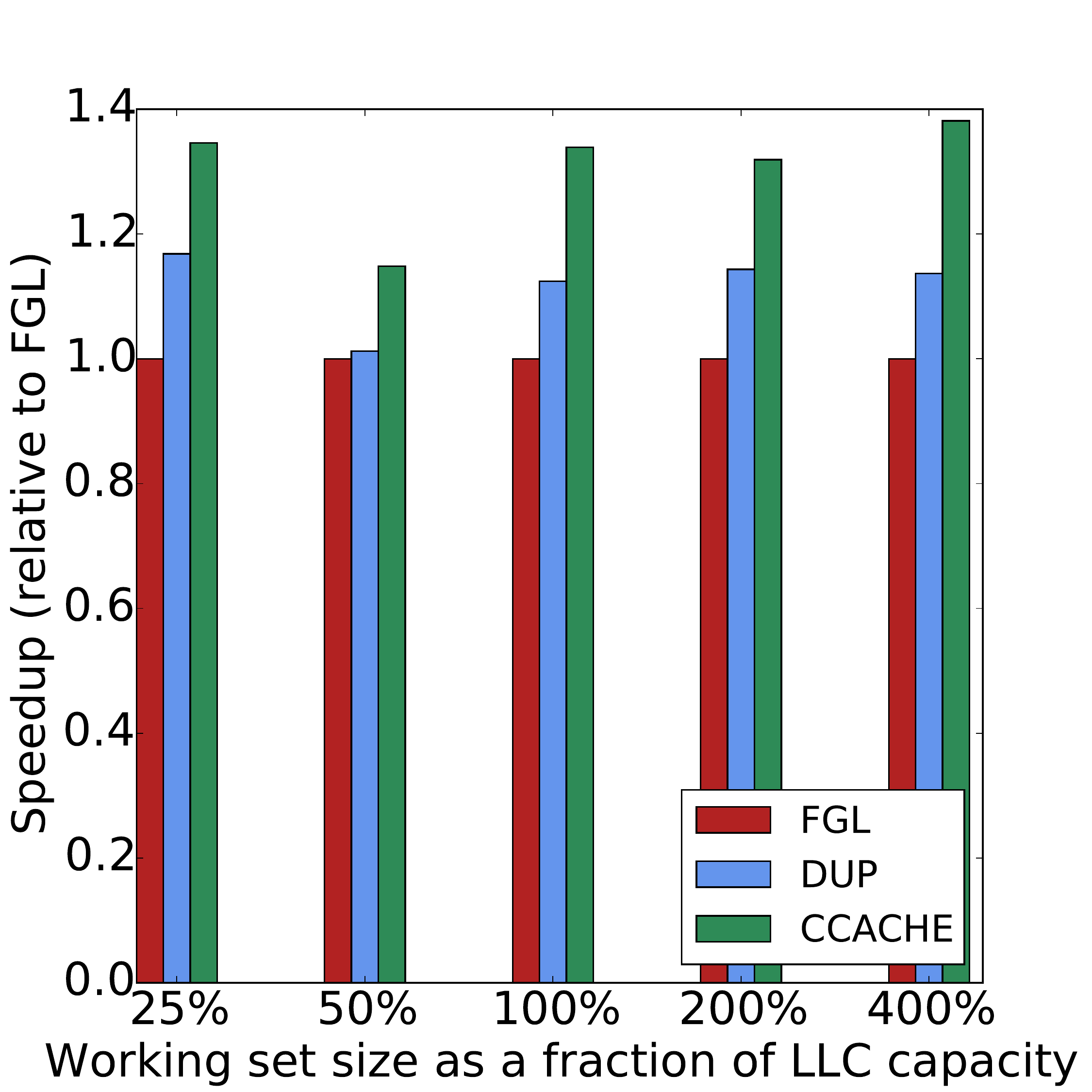}}
   & \subfloat[BFS (Kronecker)]{\label{bc-kron}\includegraphics[keepaspectratio,width=0.24\textwidth]{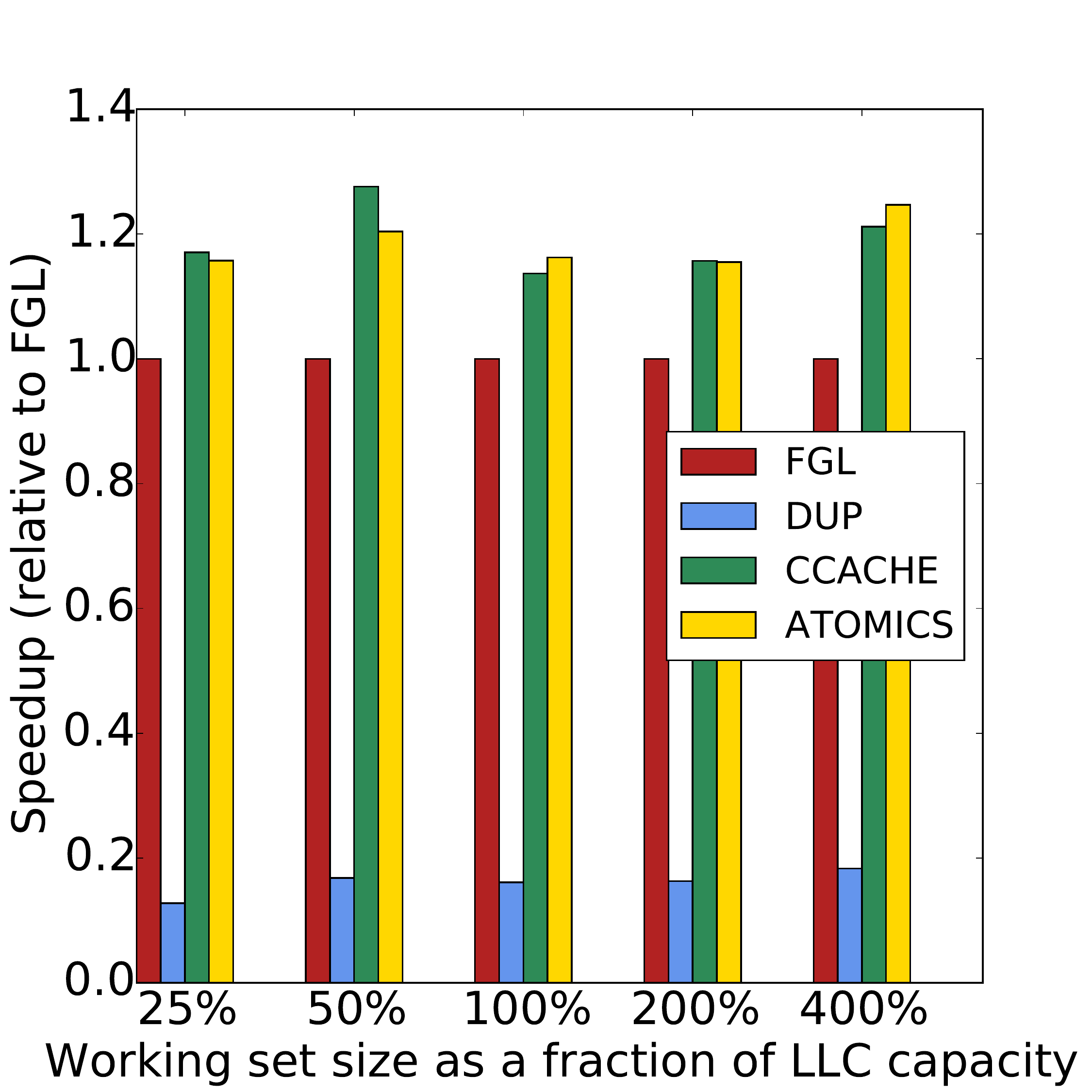}}
   & \subfloat[BFS (RMAT)]{\label{SSSP-Cycles}\includegraphics[keepaspectratio,width=0.24\textwidth]{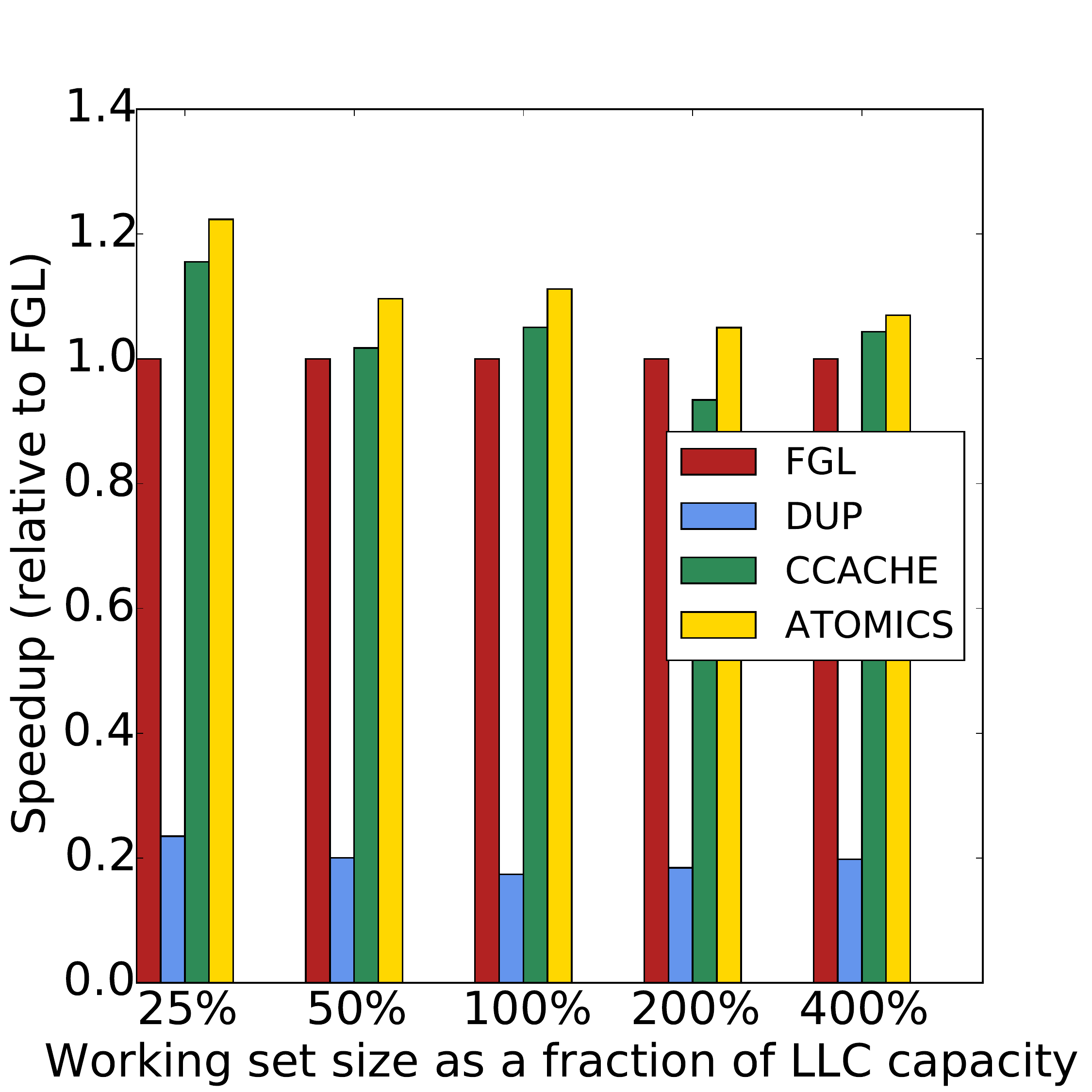}}
\end{tabular} 
\end{minipage}\par\medskip 
\centering
\begin{tabular}{ccc}
\subfloat[Key-Value Store]{\label{kv-store}\includegraphics[keepaspectratio,width=0.24\textwidth]{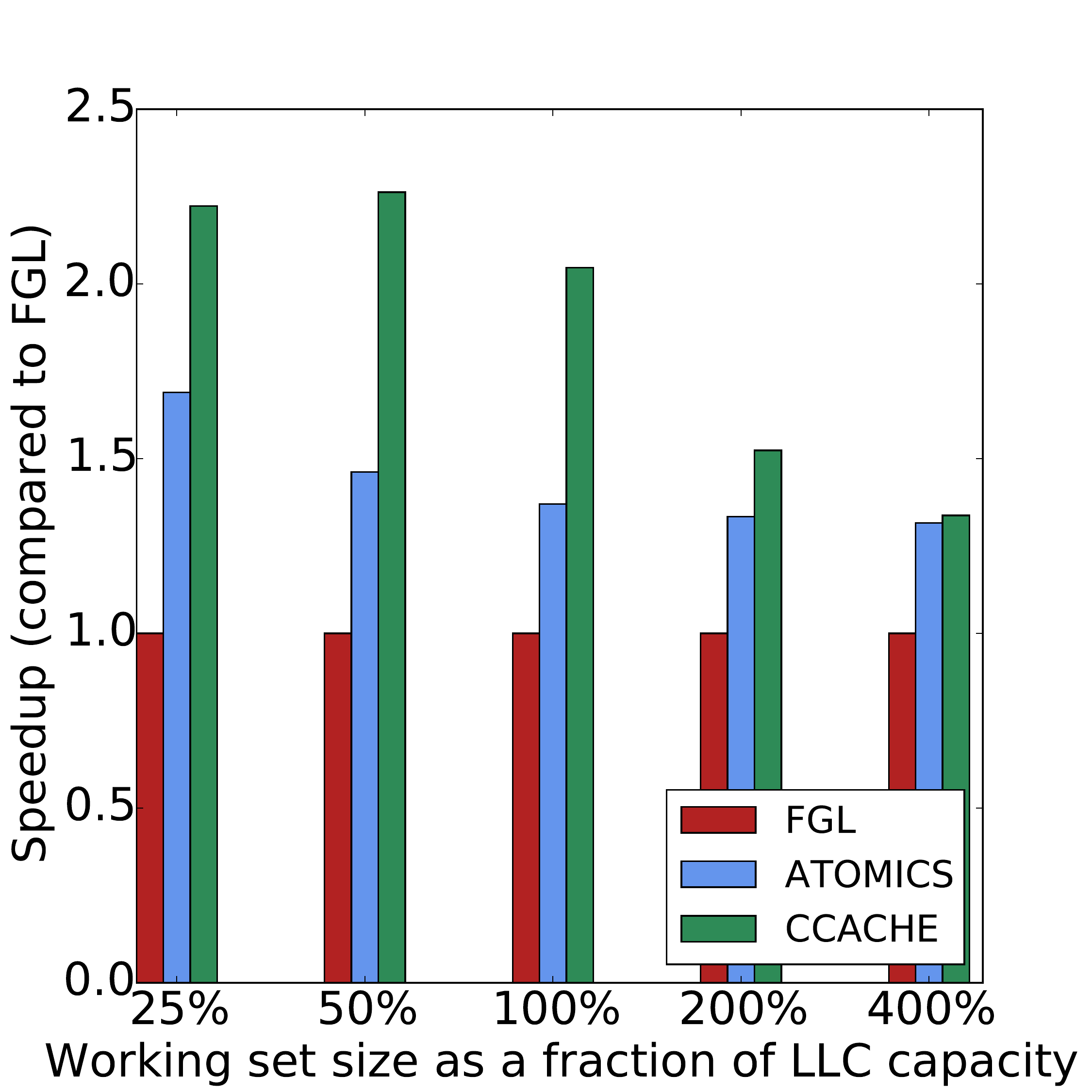}}
   & \subfloat[Page Rank (SSCA)]{\label{pr-ssca2}\includegraphics[keepaspectratio,width=0.24\textwidth]{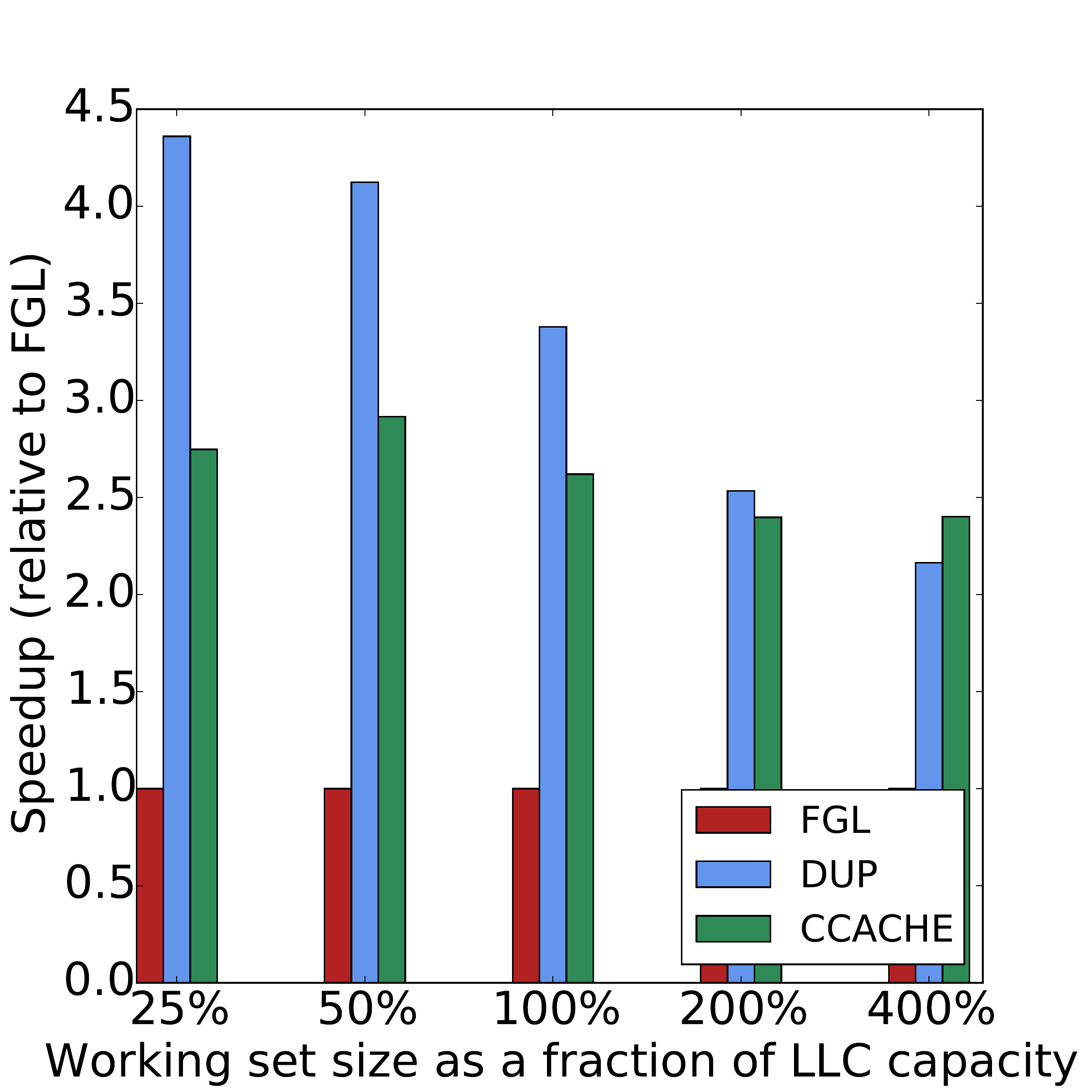}}
   & \subfloat[Page Rank (Random)]{\label{pr-random}\includegraphics[keepaspectratio,width=0.24\textwidth]{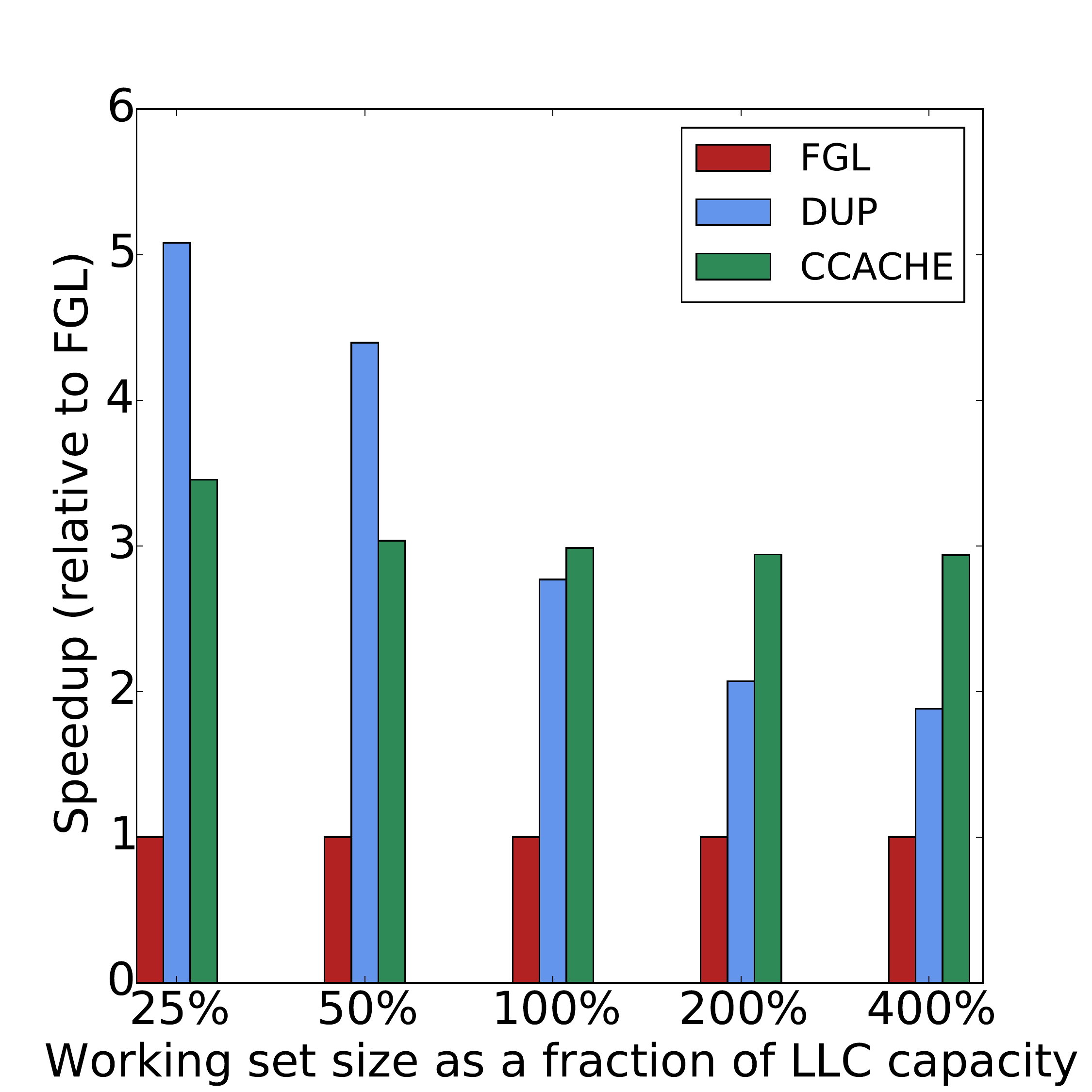}} \\
\end{tabular}  


\caption{\label{fig:perf}{\bf Performance comparison of \sys and \dup relative to \fgl.}}
\end{figure*}


We evaluated \sys to show that it improves performance compared to fine-grained locking
(\fgl) and data duplication (\dup) scalably across working set sizes. We show that \sys
improves performance compared to static duplication even with fewer hardware resources.
We also characterize our results to show they are robust to architectural parameters.


%

\subsection{Performance}

Figure~\ref{fig:perf} shows \sys's performance improvement for each benchmark
compared to \dup and \fgl for an 8 core system.  Our key result is that \sys
improves performance by upto 3.2x compared to \fgl and by upto 10x compared to
\dup across all benchmarks.  To characterize how our performance results vary
with input size, we experimented with inputs ranging from 25\% of the L3 cache
size up to 400\% of the L3 size. We report the performance improvement of \dup
and \sys versions relative to the \fgl version at each input size. 

\begin{figure}[h!]
\centering
\includegraphics[keepaspectratio,width=0.8\columnwidth]{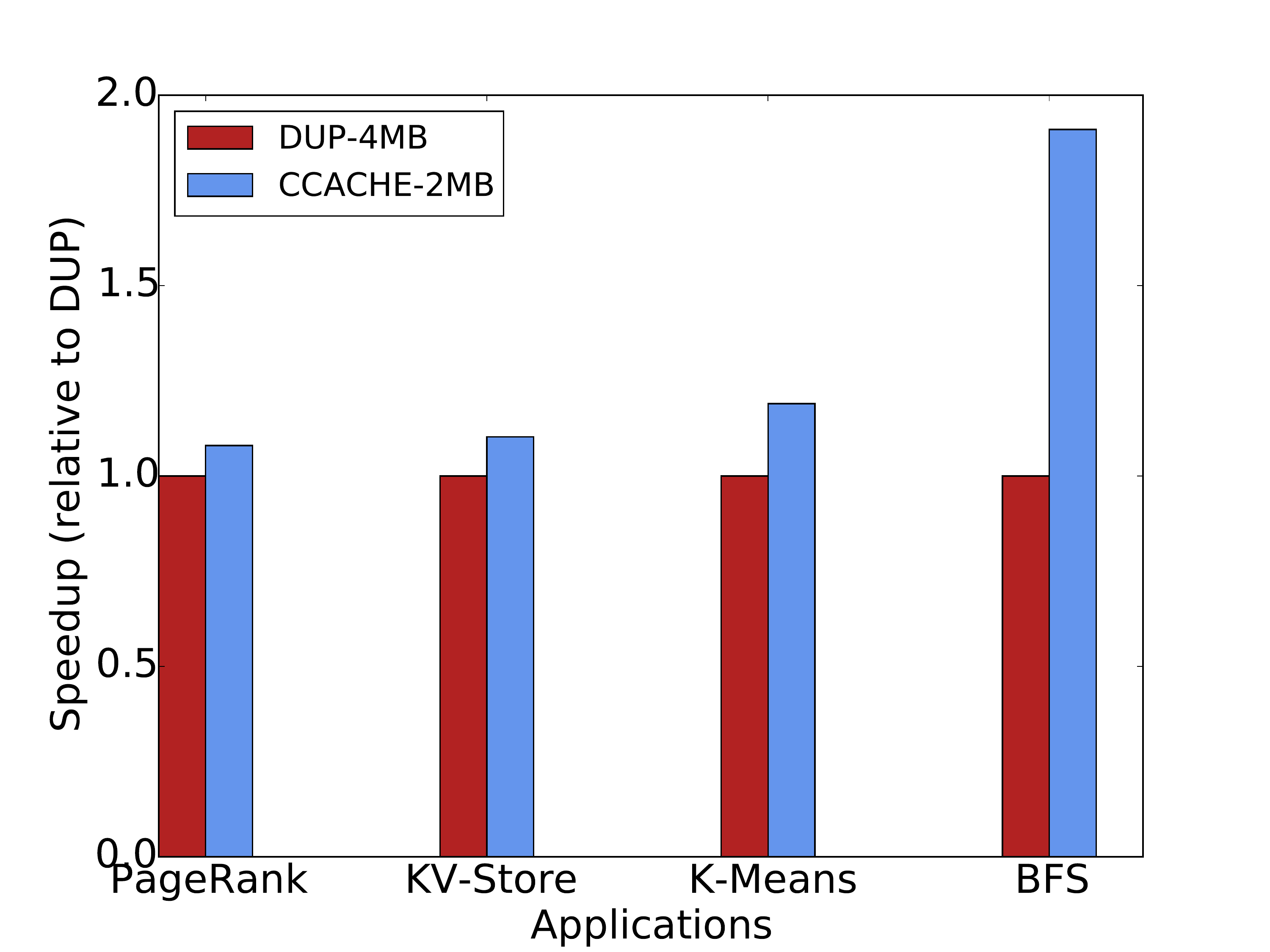}
\caption{\label{l3} {\bf \sys outperforms \dup with 50\% of L3.}}

\end{figure}

\sys hits the L3 capacity at a larger input size than \fgl (which stores locks
with data in the L3) and \dup (which stores duplicated data in the L3) because
\sys's {\em on-demand duplication} improves L3 utilization.  We evaluated the
improvement by cutting \sys's L3 capacity {\em in half } (i.e., giving \sys a
2MB L3) and comparing its performance to \dup with a full sized L3 (i.e., \dup
has 4MB of L3).  Figure \ref{l3} compares \sys's and \dup's performance when 
run on an input matching the LLC capacity. \sys is able
to provide performance improvements ranging from ~1.1X for Pagerank and KV-Store,
1.19X for K-Means, to 1.91X for BFS with half the L3 cache size. \sys on-demand
duplication is a marked improvement over \dup. 

Table~\ref{table:memory_overhead} shows the peak memory overhead of different
versions of our benchmarks when run on an input with working set size equal to  
LLC capacity. We used a mixture of
static and dynamic reasoning to estimate the maximum amount of memory a
version might use. The memory overhead for \fgl seems to be consistently
the largest for all benchmarks. This is because of the overhead of storing
fine grained locks which results in more memory than the statically duplicating
the data structure across different threads. However, in practice, we observed
that the \fgl version had fewer L3 misses than the \dup version since most of 
benchmark had significant sharing and, hence, didn't incur the peak overhead
of \fgl. The data shows that the low memory overhead of \sys helps improve 
performance compared to \fgl and \dup.

\begin{table}[h] \centering
\scriptsize
\begin{tabular}{ l | l | l | l }
\Xhline{2pt}
\textbf{App}  & \textbf{FGL} & \textbf{DUP} & \textbf{CCACHE} \\
\Xhline{1pt}
\textbf{Key-Value Store} & 12X & 8X & 1X \\
\hline
\textbf{Page Rank} & 1.91X & 1.09X & 1X \\
\hline
\textbf{K-Means} & 1X & 1X & 1X \\
\hline
\textbf{BFS} & 5.2X & 4.9X & 1X \\
%
\end{tabular}
\caption{\label{table:memory_overhead}{\bf Memory Overhead of \fgl and \dup normalized to \sys}}
\end{table}

%
 


\subsection{Characterization}
We collected additional data to characterize the performance difference between
\sys, \fgl, and \dup.  The data suggest that reductions in invalidation traffic and L3 misses
contributed to the systems' performance differences.

{\noindent \bf \dup vs. \fgl} Our performance results show that \sys consistently outperforms
the \fgl and \dup versions at larger working set sizes. However, the performance
of \fgl and \dup does not show a consistent trend across applications. For Page Rank, 
Key-Value Store and K-Means \dup outperforms \fgl by eliminating serialization 
caused by fine-grained locking and coherence traffic generated by exchange of critical
sections. In BFS, \dup's performance suffers because of the overhead of duplicating data
across different cores. These results illustrate the tensions between serialization and
coherence traffic incurred by fine-grained locking and the increase in memory footprint
by data duplication.

\begin{figure*}[tbh]
\begin{tabular}{cccc}
\subfloat[Page Rank]{\label{pr-dirmsg}\includegraphics[keepaspectratio,width=0.24\textwidth]{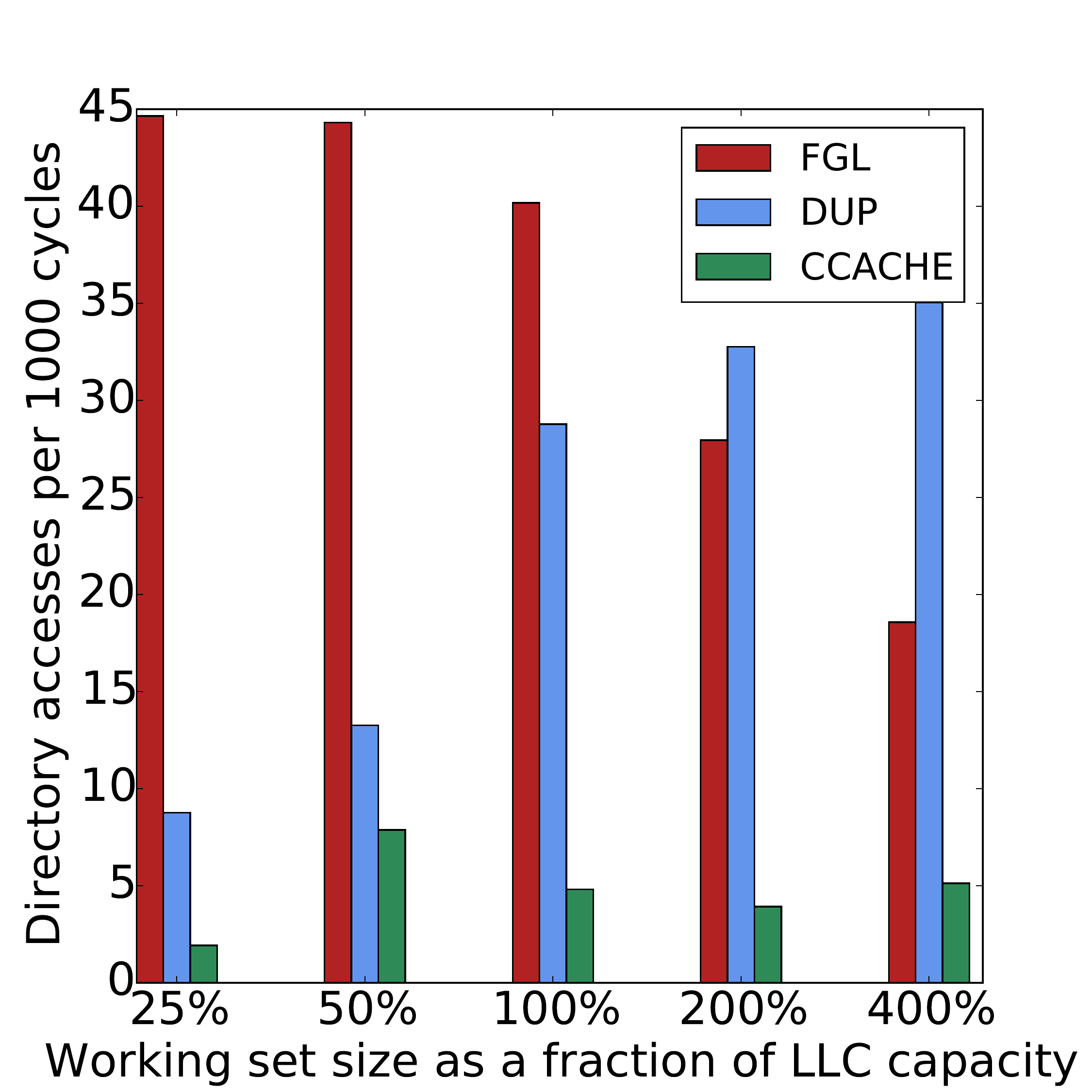}}
& \subfloat[Key-Value Store]{\label{gups-l3miss}\includegraphics[keepaspectratio,width=0.24\textwidth]{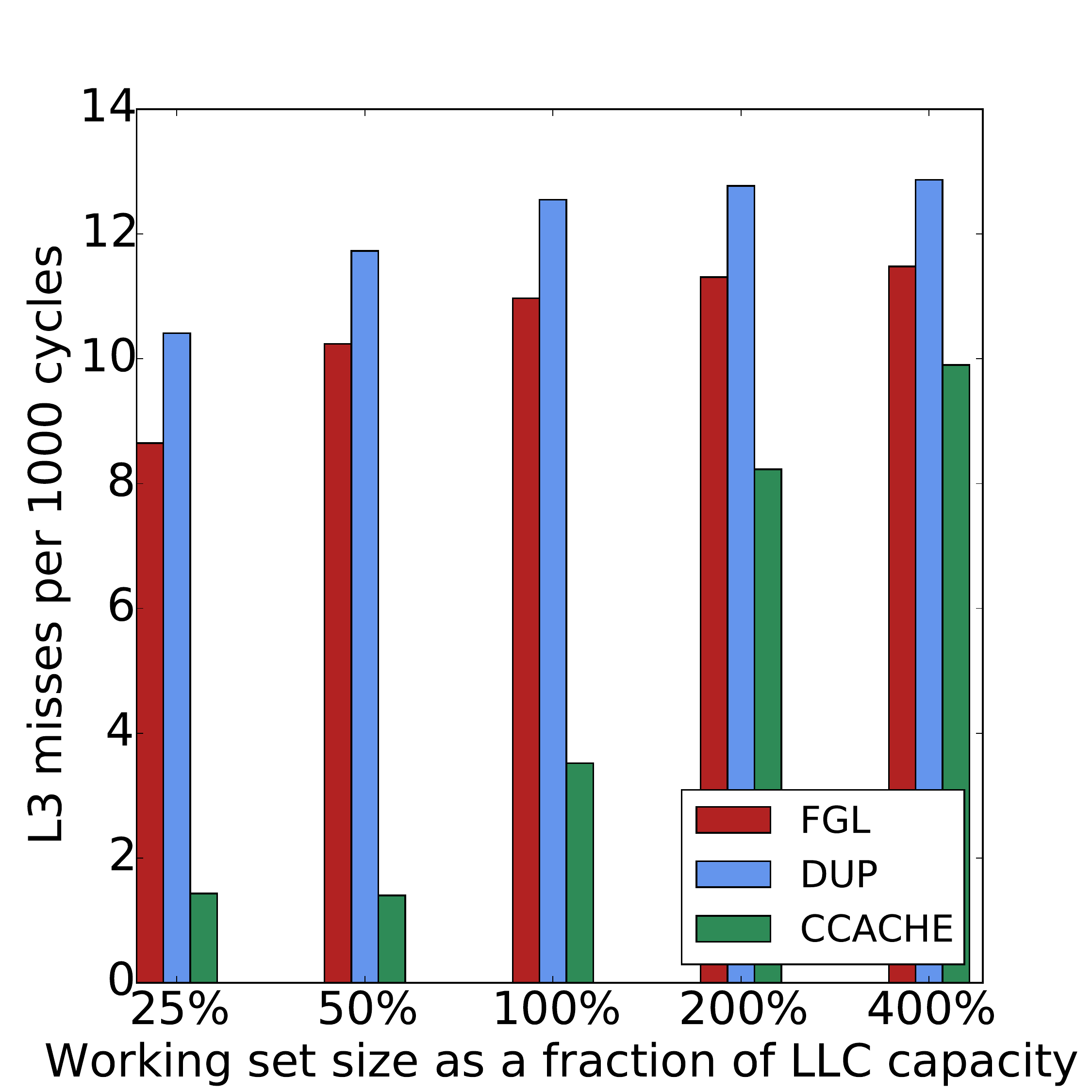}} 
& \subfloat[BFS]{\label{bfs-inv}\includegraphics[keepaspectratio,width=0.24\textwidth]{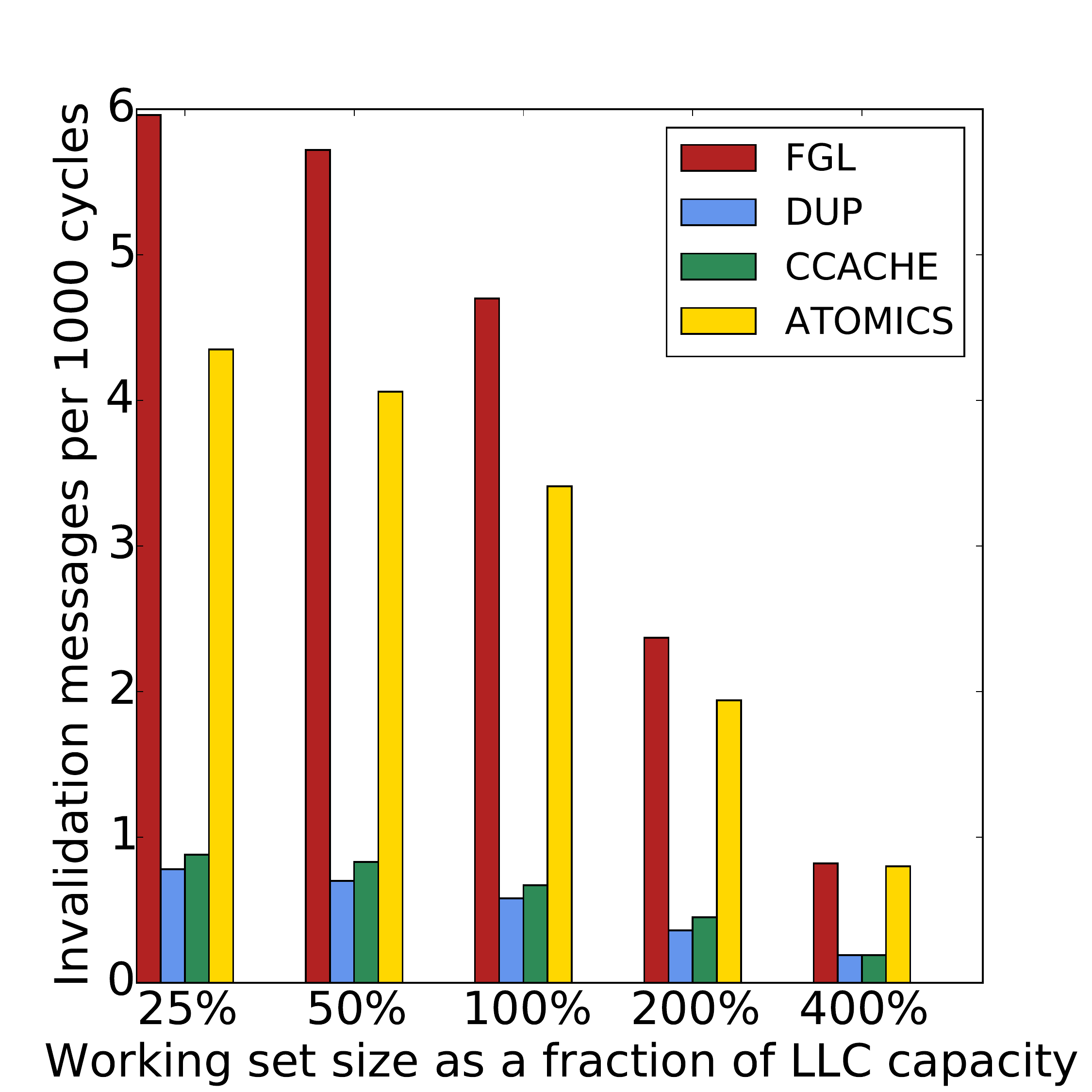}}
& \subfloat[K-Means]{\label{kmeans-inv}\includegraphics[keepaspectratio,width=0.24\textwidth]{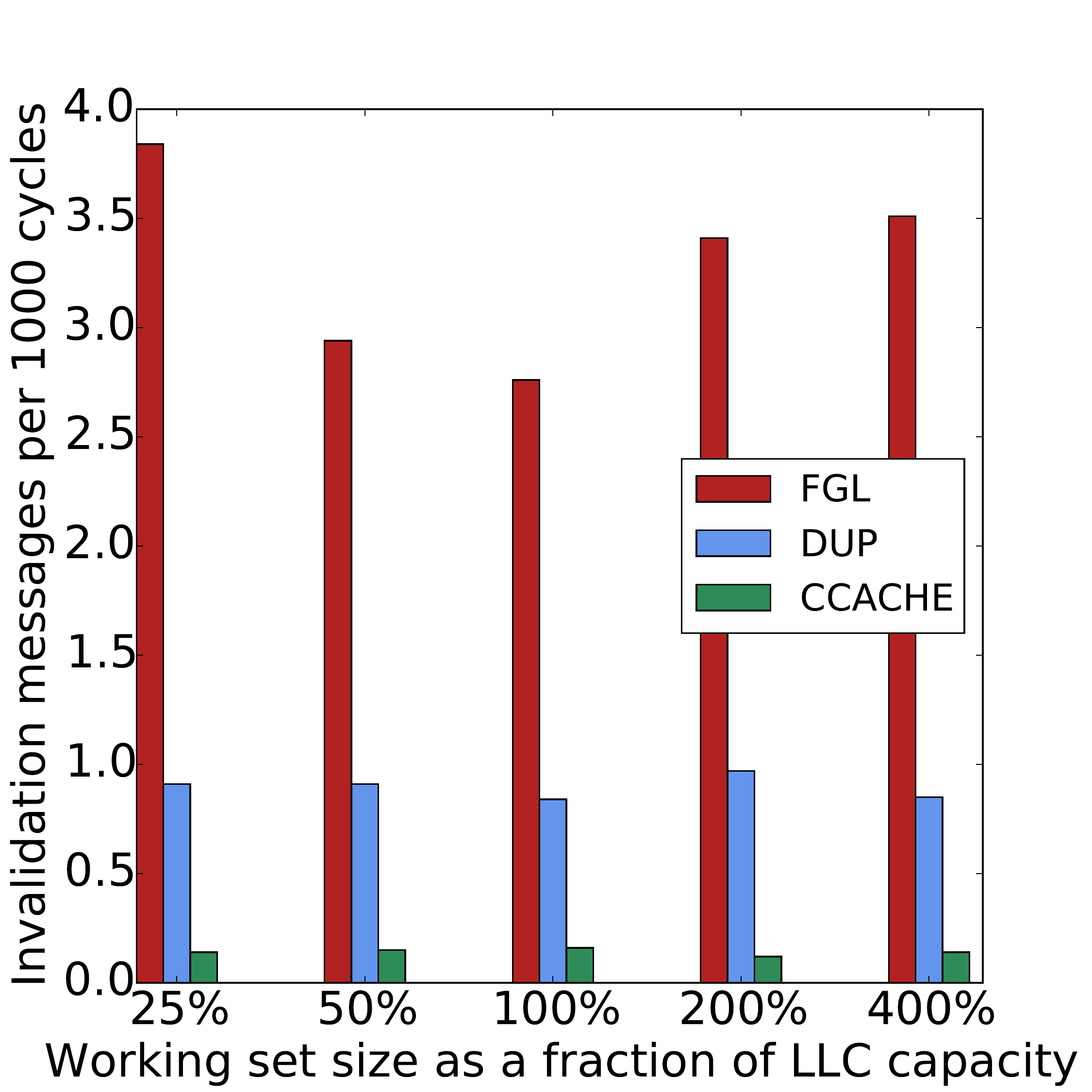}} 
\end{tabular}
\caption{{\bf Characterization.} (a) Normalized Directory accesses for Page Rank (b) Normalized L3 cache misses for Key-Value Store and Normalized Invalidation messages for (c) BFS and (d) K-Means. } 
\end{figure*}

{\noindent \bf Page Rank.} Figure~\ref{pr-dirmsg} shows the number of
directory messages issued per 1000 cycles for our three versions when run on
the random graph input. The reduction in directory accesses in \sys compared 
to the \fgl and \dup versions explains the speedup achieved by \sys. The 
Further, the increase in directory accesses of \dup with larger working sets
corresponds with the reduced performance improvement provided by \dup for 
large working sets. We also observed a decrease in the number of L3 misses
incurred by \sys compared to \dup and \fgl, which could also contribute to 
\sys performance improvement. Note that \sys was able to outperform our
highly optimized \dup implementation for larger, more realistic working set 
sizes without imposing the burden of efficient duplication on the programmer.



{\noindent \bf Key-value Store.} Figure \ref{gups-l3miss} shows the fraction of
L3 misses per 1000 cycles for \fgl, \dup and \sys. The main result is that \sys's
performance improvement for \gups corresponds to the reduction in L3 misses. The
data also shows that \sys incur fewer L3 misses (2.5--3X fewer) than \dup and 
\fgl when the working set size matches LLC capacity, further illustrating that
\sys better utilizes the LLC. We also observed a consistent reduction in the 
number of invalidation signals issued in \sys compared to \fgl. The reduction
in invalidation traffic also likely contributes to \sys's 2.3X performance 
improvement.

{\noindent \bf BFS} Figure \ref{bfs-inv} shows the invalidation traffic per 1000
cycles for \fgl, \dup, \sys and atomics versions. The result shows a significant
reduction in invalidation traffic in the \dup and \sys versions compared to \fgl
and atomics versions. The difference in the normalized invalidation traffic 
across different working set sizes corresponds with the performance improvement of
\sys over \fgl and atomics. We also observed that \sys and atomics incurred about
the same number of L3 misses which was substantially fewer than those incurred by
\fgl and \dup. This could explain the bigger performance 
gap of \sys over \fgl and \dup compared to atomics. \sys provides the performance 
benefits of atomic instructions while being more generally applicable to a variety
of commutative updates. We discuss \sys's generality in more detail in Section~\ref{sec:generality}


{\noindent \bf K-Means.} Figure \ref{kmeans-inv} shows the invalidation  
traffic normalized to cycle count for the three versions when run on the 
floating point dataset, illustrating the likely root of \sys's performance
improvement for K-Means.  \sys has less coherence traffic than \fgl because
\sys operates on private duplicates when \fgl must synchronizes and make
updates globally visible.  \fgl requires coherence actions to keep both locks
{\em and} data coherent, which \sys need not do.  \sys also had fewer coherence
actions than \dup for K-Means because \sys's merge differs from \dup's. During
a \dup merge, one thread iterates over all cores' copies of the data, to
produce a consistent result.  The merging core incurs a coherence overhead to
invalidate the duplicated data in every other core's cache.  After the merge,
each core that had its duplicate invalidated must re-cache the data, incurring
yet more coherence overhead.  \sys cores avoid the coherence overhead by
manipulating data in their L1s and merging {\em their own} data. 


\subsection{Support for Diverse Merge Functions}\label{sec:generality}
To demonstrate \sys's flexibility in supporting arbitrary merge functions, we 
implemented a saturating counter and complex number multiplication version of
Key-Value Store and an approximate merge version of kmeans. In the saturating
counter benchmark, \sys's merge function reduces privatized copies up to 
a threshold. For complex multiplication, 
the merge function complex-multiplies privatized copies.
We showed that \sys also flexibly supports
approximate computing by writing an approximate merge function for K-Means.
The approximate merge version discards updates for some points in a dataset, which does not significantly
alter cluster centers. We randomly
dropped 10\% of a core's merge operation, which leads to 20\% degradation
in the intra-cluster distance metric. In some cases, quality 
degradation is tolerable and \sys allows the programmer to make
such a quality-performance trade-offs. Our evaluation showed that \sys 
provides a speedup over \fgl and \dup similar to the 
baseline versions of these three applications (Figure~\ref{fig:perf}).
The results show that \sys's performance benefits are not restricted only
to applications with only simple commutative operations and can be
extended to arbitrary commutative updates.

\subsection{Characterizing the Merge-on-evict
Optimization}\label{subsec:soft-merge}

By default \sys uses the merge-on-evict and dirty-merge optimizations 
(Section~\ref{subsec:optimizations}). We studied the benefit of these 
optimizations by re-running our benchmarks without the optimizations and 
comparing to the performance with the optimization. Both optimizations 
did not significantly improve the performance of un-optimized \sys 
because merge functions comprise only a small fractions of total cycles
executed for all our benchmarks.

\begin{figure}[h]
\centering
\includegraphics[keepaspectratio,width=0.7\columnwidth]{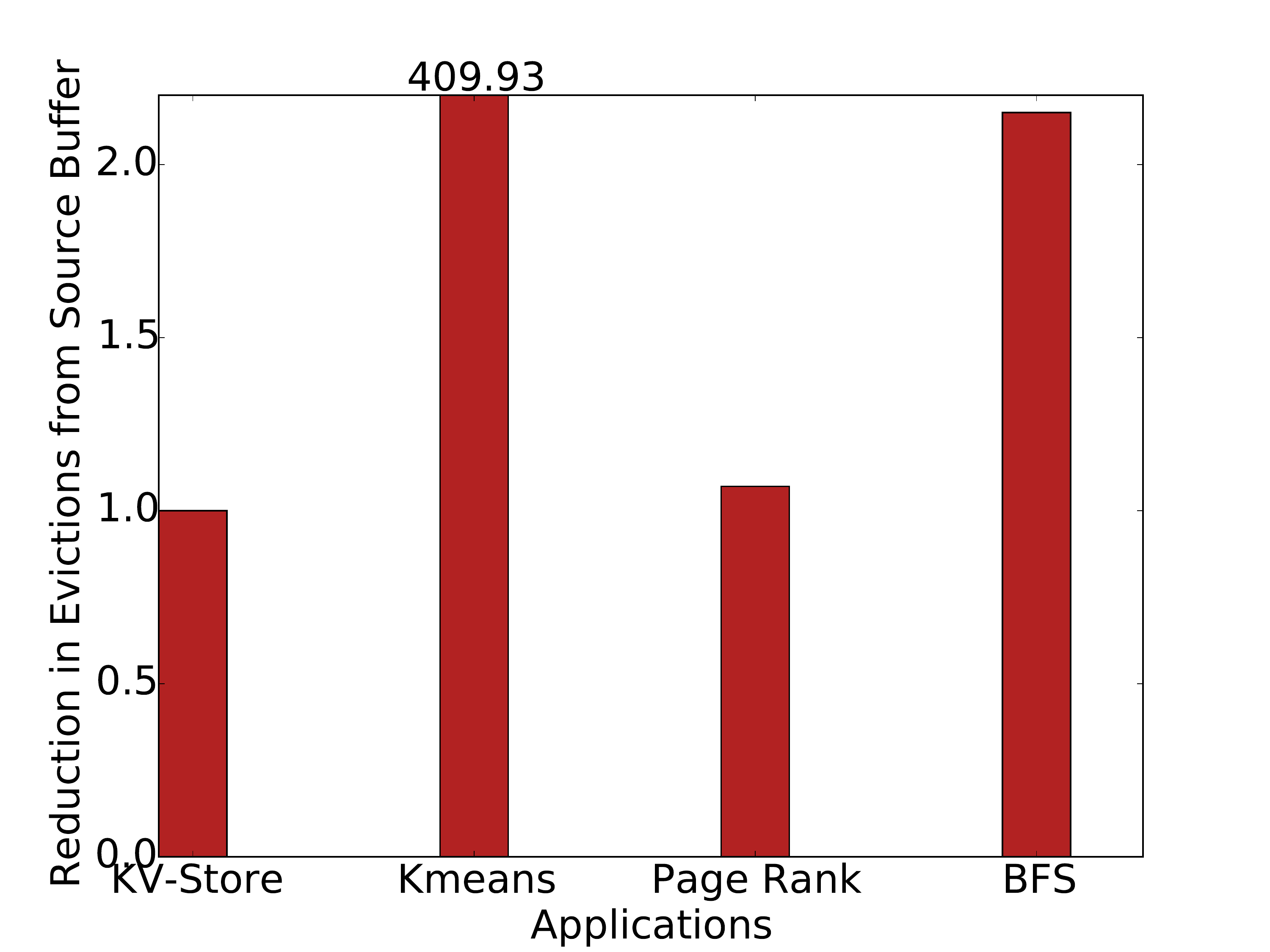}
\caption{\label{fig:lazy_merge} {\bf Merge-on-Evict reduces Source Buffer Evictions}}
\end{figure}

While the optimizations did not improve the performance of a baseline \sys
implementation, they are essential for improving performance over \dup and
\fgl versions. The merge-on-evict optimization improves locality at the 
source buffer, requiring fewer merges and, consequently, reduced locking of LLC 
lines. Figure~\ref{fig:lazy_merge} shows the reduction in source buffer 
evictions by out merge-on-evict optimization compared to a \sys 
implementation without the optimization. The optimization
reduced the number of evictions by 2.2X in BFS and 409.9X in K-Means. The 
increased source buffer locality makes \sys a more efficient alternative 
to data duplication than \dup. We also evaluated the performance benefits 
of the dirty-merge optimization. The optimization reduces the number 
of merges required by only merging data updated by a core. Our evaluation
showed that this optimization does not provide performance benefit to 
update heavy benchmarks like K-Means, Key-Value store and BFS. However, 
for Page Rank, where a lot of \cdata is only read and never updated, the 
dirty merge optimization was able to reduce the number of merges performed
by 24X compared to a \sys implementation without the optimization

\section{Related Work} 

Several areas of prior work relate to \sys.  Section~\ref{sec:background}
discussed explicit data duplication and reductions. This section discusses
recent work on COUP~\cite{coup} and then discuss work on: (1) combining
parallel updates; (2) {\em expansion} and {\em privatization}; and (3) {\em
speculation}, including transactions.

The closest prior work to \sys is COUP\cite{coup}, which uses commutativity to
reduce data movement. COUP extends the coherence protocol to support
commutativity and supports a small, fixed set of operations in hardware. While
similar, \sys differs significantly.  \sys is more flexible, allowing
programmer-defined, software commutative operations. In contrast, COUP supports
a fixed set of hardware commutative operations only. This key difference makes
\sys more flexible and broadly applicable than COUP.  Section~\ref{sec:eval}
evaluates \sys's flexibility with a spectrum of merge types.  Additionally,
COUP requires coherence protocol changes and \sys does not.  In \sys, \cops
never generate outgoing coherence requests, and are never the subject of
incoming requests; \cdata lines need no coherence actions and non-\cdata remain
coherent as usual.  Lastly, COUP cannot exploit \sys's merge-on-evict optimization 
because COUP does not get information from the programmer (i.e., \smrg vs. \hmrg).

\subsection{Combining Independent Parallel Updates}

Prior systems supported combining the result of {\em independent} executions of
{\em dependent} operations.  Parallel prefix~\cite{parallelprefix} computations
broke dependences by recursively decomposing operations and later combining
partial results, similar to how \sys merges updates.  

Commutativity analysis~\cite{commutativityanalysis} identifies and parallelizes
commutative regions of code, inserting code to combine commutative partial
results.  \sys draws inspiration from this work, but differs considerably,
allowing arbitrary, programmer-defined merge functions and targeting hardware.

Commutative set~\cite{commutativesetPLDI2011} is a language that allows
specifying commutative code blocks. A compiler can then produce code that
executes commutative blocks in parallel, serializing them on completion.  The
main distinction from \sys is that \sys's parallelization is on-demand and
avoids the need for compiler analysis by using hardware support.

Concurrent revisions~\cite{revisionsOOPSLA10, revisionsOOPSLA11} followed the
tack of commutativity analysis, promoting an execution model that allows a
software thread to operate on a copy of a shared data structure.  The central
metaphor of this work is ``memory as version control''.  The system resolves
conflicting parallel updates with a custom merge function.  This work's
execution model was motivational for \sys's use of duplication and merging.  A
key difference is that \sys uses architecture, requiring very few software or
compiler changes. 

RetCon~\cite{retconISCA10} operates on {\em speculative} parallel copies of
data in transactions.  When transactions conflict, RetCon avoids rollback by
{\em symbolically tracking} updated values.  Applying an update derived from
symbolic tracking is like \sys's use of a merge function to combine partial
results.  What differs is that symbolic tracking is limited in the types of
merges it can perform.  RetCon cannot perform merges that cannot be represented
with the supported, symbolic, arithmetic expressions.  \sys's permits general
merge functions and does  not incur the cost of speculation.

Symple~\cite{sympleSOSP15} automatically parallelizes dependent operations to
user-defined data aggregations, also using symbolic expressions.    Symple
treats unresolved dependences as symbolic expressions, eventually resolving
them to concrete values.  Like Symple, \sys allows manipulating shared data and
merging partial results.  \sys differs in its use of architecture and lack of
need for symbolic anlaysis, which is likely to be complex in hardware.

\subsection{Duplication, Expansion, and Privatization}

Other techniques looked at automatic software parallelization using {\em
expansion}~\cite{expansionCACM86,expansionPLDI13}, {\em data duplication and
reductions}~\cite{mapreduceOSDI04,mapreduceCACM08,openmpreduce,cilkhyperobjects},
and privatization~\cite{arrayprivatizationLCPC94}.

Expansion makes copies of scalars~\cite{expansionCACM86}, data
structures~\cite{expansionPLDI13}, and arrays~\cite{arrayprivatizationLCPC94},
allowing threads to manipulate independent replicas.
Expansion, especially of large structures, is like 
duplication in our evaluation~\ref{sec:eval}.  Expansion 
risks excessive cache pressure, especially in the single-machine,
in-memory workloads that we target.

Data duplication and reduction has wide-spread adoption in parallel
frameworks~\cite{mapreduceOSDI04,mapreduceCACM08,openmpreduce,cilkhyperobjects}.
These systems focus on scaling to big data and large numbers of machines,
unlike  \sys, which does not require a language framework, instead leveraging
hardware to avoid static duplication.

Copy-on-Write (CoW) techniques~\cite{coredet,determinator,grace} privatize
data, duplicating at updates. \sys differs considerably, not requiring
allocation or memory remapping for copies.  Furthermore, \sys supports
arbitrary merging, instead of ordered, last-write-wins updating.

\subsection{Speculative Privatization}

A class of techniques use {\em speculation} to parallelize accesses to shared
data.  Speculation increases parallelism, but has high software overheads and
hardware complexity.

Software~\cite{stmTOCS07} and hardware transactions~\cite{htmHPCA05} buffer
updates (or log values) and threads compute independently on shared data.
Mis-speculation aborts and rolls back updates, re-executing to find a
conflict-free serialization.  The similarity to \sys is that transactional
threads manipulate isolated values.  However, transactions abort work on a
conflict, rather than trying to produce a valid serialization. By contrast,
\sys's merge function aggressively finds a serialization, despite conflicts.

Both speculative multithreading~\cite{tlsASPLOS98,tlsISCA00,stampedeTOCS05}
and bulk speculative
consistency~\cite{tccASPLOS04,bulkscISCA07,storewaitfreeISCA07,invisifenceISCA09}
are transaction-like execution models that {\em continuously} dynamically
duplicate data, enabling different threads to operate on duplicates in
parallel.  Like most other transactions work, these efforts
primarily {\em roll back} work when the system detects an access to the same
data in different threads.  In contrast, \sys merges manipulations of
shared data in different threads.  

Prior work on TMESI~\cite{tmesi} also modified the coherence protocol to
support programmable privatization for transactions.  \sys also offers a form
of programmable privatization, but differs in several ways. 
\sys does not require a large number of additional coherence protocol states to
handle privatized data.  \sys has only a single ``state'' -- the \sys bit --
because it privatizes commutatively updated data only, and those data are kept
coherent by merging. 
Unlike TMESI, \sys avoids the cost of speculation, providing atomicity at the
granularity of cache lines only, not transactional read and write sets.
Moreover, \sys is applicable to lock based code, while TMESI is specific to
transactions.

\section{Conclusions and Future Work}

We presented \sys, a system that improves the performance of parallel programs
using {\em on-demand data duplication}.  \sys improves the performance of
accesses to memory that are {\em commutative}.  Leveraging the fact that
commutative operations can execute correctly in any order, \sys allows each
core to operate on involved data independently, without coherence actions or
synchronization.  Merging combines cores' independently computed results with
memory, producing a consistent, coherent final memory state.  Our evaluation
showed that \sys considerably improves the performance of several important
applications, including clustering, graph processing, and key-value lookups,
even earning a performance improvement over a system with twice the amount of
L3 cache.  The future for \sys goes in two directions. First, leveraging other
high-level properties, such as approximability, to extend its benefits to
programs with non-commutative operations.  Second, we envision \sys-like
support to remediate conflicts between commutative operations in
conflict-checking parallel execution
models~\cite{conflictexceptions,valor,drfx}.

\bibliographystyle{abbrv}
\bibliography{ref}

\end{document}